\documentclass[aps,prc,twocolumn,superscriptaddress]{revtex4-2}
\usepackage{lineno,isotope,mathrsfs}
\modulolinenumbers[5]
\usepackage{amsmath,bbold,amssymb,epsfig,bm,mathrsfs,feynmp}
\usepackage{color,slashed,exscale,multirow,soul,nicefrac,gensymb}
\usepackage{longtable,times,txfonts,xcolor,graphicx,tikz}
\usepackage[normalem]{ulem}
\usepackage[breaklinks,colorlinks,citecolor=blue]{hyperref}
\definecolor{red}{rgb}{0.8,0,0}
\definecolor{violet}{rgb}{0.4,0,0.4}
\definecolor{green}{rgb}{0,0.5,0.0}
\definecolor{navy}{rgb}{0.0,0.0,0.6}
\definecolor{orange}{rgb}{0.8,0.2,0.0}
\usepackage[normalem]{ulem}  % \sout{old text} for strikeout  

\newcommand{\bea}{\begin{eqnarray}}
\newcommand{\eea}{\end{eqnarray}}

\newcommand{\MR}{$M$-$R\ $}

\begin{document}
\title{
Bayesian inferences on covariant density functionals 
from multimessenger astrophysical data: \\
The impacts of likelihood functions of low density 
matter constraints
}
%
%\author[0000-0001-8635-3939]{Jia-Jie Li}
\author{Jia-Jie Li}
\email{jiajieli@swu.edu.cn}           
\affiliation{School of Physical Science and Technology, 
             Southwest University, Chongqing 400715, China}        
%
%\author[0000-0001-9626-2643]{Armen Sedrakian}
\author{Armen Sedrakian}   
\email{sedrakian@fias.uni-frankfurt.de}         
\affiliation{Frankfurt Institute for Advanced Studies,
             D-60438 Frankfurt am Main, Germany}         
\affiliation{Institute of Theoretical Physics,
             University of Wroc\l{}aw, 50-204 Wroc\l{}aw, Poland} 
\begin{abstract}
We systematically investigate how the choice between Gaussian 
and uniform likelihood functions in Bayesian inference affects 
the inferred bulk properties of compact stars and nuclear matter 
within covariant density functional-based equations of state. 
To enable direct comparison between the two approaches, we designed 
the uniform likelihood function with a Gaussian-equivalent 
normalization factor and marginalization behavior. Across three 
representative astrophysical scenarios, both approaches yield nearly 
identical mass-radius relations, density-pressure relations, and 
overlapping 95.4\% confidence level regions. 
Although our inference analysis is carried out using 
parameters of the density functional, we subsequently determine 
the associated nuclear matter characteristic coefficients derived 
from the Taylor expansion of the energy density around the saturation 
density. We observe significant variation in the predicted isoscalar 
channel coefficients (e.g., the nuclear incompressibility) across 
different astrophysical scenarios, while the isovector channel 
(e.g., the slope of symmetry energy) exhibits only minimal variation.

\end{abstract}
%
%\keywords{Neutron stars (1108); Neutron star cores (1107); 
%Nuclear astrophysics (1129); Gravitational waves astronomy (675)}
%
\date{\today}
\maketitle
%
%----------------------------------------------------------
\section{Introduction}
\label{sec:Intro}
%----------------------------------------------------------

Understanding the physics of nuclear matter over a wide range 
of densities and isospin asymmetries is essential for the description 
of the behavior of neutron-rich nuclei, heavy-ion reactions, 
the structure of compact stars (CSs), and related problems. 
In particular, the equation of state (EOS) of nuclear matter 
can be derived from modeling the physics of these systems within 
different approaches, ranging from density functional methods 
to microscopic {\it ab initio} calculations. In the process of 
discerning the valid set of the EOS, the statistical inference 
of its parameters from the data, including multimessenger 
astrophysics, is of crucial importance.

The Bayesian inference framework, involving various theoretical 
and observational constraints, has been applied to EOS models 
covering the {\it ab initio} calculations combining model-agnostic 
extrapolations such as the use of polytropes or speed-of-sound 
schemes~\citep{Raaijmakers:2019,Landry:2020,Raaijmakers:2020,
Legred:2021,Pang:2021,Altiparmak:2022,Annala:2022,Annala:2023,
Chimanski:2022,Rutherford:2024,Fan:2024},
meta-type models parametrized by nuclear characteristic 
parameters~\citep{Margueron:2018b,Margueron:2019,Zhang:2020,Tsang:2024},
non-relativistic models based on effective nuclear potentials 
like the Skyrme interactions~\citep{Mondal:2023,Zhou:2023,
Beznogov:2024a,Beznogov:2024b,Tsang:2024},
and covariant density functional (CDF) based
models~\citep{Traversi:2020,Malik:2022a,Malik:2022b,
Zhu:2023,Beznogov:2023,Malik:2023,Salinas:2023,
Providencia:2023,Salinas:2023,Char:2023,Huang:2024,
Scurto:2024,Lijj:2025a,Char:2025,Lijj:2025b}.

The CDF models are commonly separated into two broad classes: 
those with non-linear meson contributions to the effective Lagrangian, 
and those that maintain only linear coupling but incorporate 
density-dependent couplings to account for medium modifications 
of meson-nucleon vertices. Both approaches offer flexibility, 
the first class through various self- and cross-couplings between 
mesons, and the second class through different forms of density-dependence. 
This adaptability allows one to tailor the models to address specific 
physics questions, see Refs.~\citep{Vretenar:2005,Niksic:2011,Oertel:2017,Piekarewicz:2020,Sedrakian:2023} for reviews.
In a CDF-based Bayesian analysis, current knowledge of the 
isospin-asymmetric EOS is typically encoded as constraints on 
the Taylor expansion coefficients of energy density of nuclear 
matter close to saturation density ($\rho_{\rm sat}$) and the 
isospin symmetric limit. Specifically, the incompressibility 
($K_{\rm sat}$) and symmetry energy slope ($L_{\rm sym}$) parameters, 
determined from giant resonance experiments or nuclear collision data, 
provide important complementary constraints. The low-density region 
can be further calibrated using microscopic computations, which are 
typically based on potentials derived from chiral effective field 
theory ($\chi$EFT) calculations, which are suitable for describing 
nuclear matter at densities 
$\le 2\,\rho_{\rm sat}$~\citep{Hebeler:2013,Lynn:2016,Drischler:2019,Huth:2021}.

In addition to the various CDF models used in Bayesian analysis, different 
approaches exist for imposing low-density matter constraints from theories 
and measurements, including their uncertainties.
The most common method involves variable likelihood functions, particularly 
the Gaussian-type function, which builds a probabilistic model accounting for 
uncertainty in each constraint.
An alternative approach is the Heaviside method, which makes no specific 
assumptions about uncertainty distributions but simply rejects EOS models 
outside defined lower and upper boundaries. Gaussian functions are typically 
used for implementing nuclear matter Taylor expansion coefficient constraints 
with preferred values, while the Heaviside method is more commonly applied 
for $\chi$EFT constraints, see for example, Refs.~\citep{Malik:2022a,Malik:2022b,Beznogov:2023,Malik:2023,
Salinas:2023,Providencia:2023,Char:2023,Scurto:2024,
Lijj:2025a,Char:2025,Lijj:2025b}.
Recently, a third approach was proposed - a 
``uniform-Gaussian combination" method~\citep{Scurto:2024}, where the $1\,\sigma$ domain of a Gaussian distribution is replaced by a uniform distribution, providing a hybrid framework for constraint implementation. This hybrid approach will be referred to as the ``uniform-Gaussian-combination".

One of the ongoing efforts in multi-messenger constraints on the EOS is to investigate the model dependence in statistical inferences. Within a given EOS framework, uncertainties in the inference naturally arise from the constraints used, their associated uncertainties, and the form of the applied likelihood functions. In the present study, we focus specifically on the sensitivity of the CDF-based Bayesian analysis to the choice of likelihood.

We employ both Gaussian and uniform-Gaussian-combination likelihood functions to characterize low-density nuclear matter properties. In doing so, we maintain the same astrophysical constraints to study how different likelihood function choices impact Bayesian analysis predictions. This work is motivated by the observation that numerous analyses consistently constrain only the first few coefficients of the Taylor expansion of nuclear matter near $\rho_{\rm sat}$ to within 5\%~\citep{Typel:1999,Lalazissis:2005,Dutra:2014,Margueron:2018b}. This is not the case for higher-order coefficients~\citep{Margueron:2018b,Sunboyang:2024}. 

For example, the incompressibility parameter $K_{\rm sat}$ has been quite well established from giant monopole resonance (GMR) experiments to be $230 \pm 40 \mathrm{MeV}$~\citep{Youngblood:1999,Todd-Rutel:2005,Shlomo:2006,Dutra:2014,Garg:2018,Litvinova:2022,Colo:2023}. Large systematic uncertainties are commonly associated with its value inferred from heavy-ion collision analyses.
Elliptic flow measurements compared with isospin quantum molecular dynamics (IQMD) simulations suggest an incompressibility of $K_{\text{sat}}=190\pm 30$ MeV~\citep{LeFevre:2016}. Proton directed and elliptic flow measurements compared with pBUU transport models indicate $K_{\text{sat}}=210$\textendash{}$300$~MeV~\citep{Danielewicz:2002}. Analyzing the Kaon yields experimental results together with RQMD and IQMD calculations enabled extraction of the EOS characterized by $K_{\text{sat}} = 200$~MeV~\citep{Fuchs:2001,Hartnack:2006}. Comparison of FOPI data for rapidity-dependent elliptic flow with UrQMD simulations gives $K_{\text{sat}}=220\pm40$~MeV~\citep{Wangyj:2018}. Recently, a Bayesian analysis of STAR flow data suggests a relatively hard EOS with $K_{\text{sat}}=285 \pm 67 \mathrm{MeV}$~\citep{Oliinychenko:2023}. This broad range of constraints on $K_{\rm sat}$ indicates that there is no consensus value yet. 

Similar challenges exist in constraining the symmetry energy slope parameter $L_{\rm sym}$~\citep{Russotto:2011,Russotto:2016,PREX:2021,CREX:2022,Reed:2021,Reinhard:2021,Essick:2021,SpiRIT:2021}. The value of $L_{\text{sym}}$, previously established to be around 60 MeV~\cite{Oertel:2017}, has been challenged by recent experimental findings. The PREX-II experiment, which measured the neutron skin thickness of ${ }^{208}$Pb, suggests a significantly higher value of $L_{\text{sym}}=106 \pm 37$~MeV~\citep{PREX:2021,Reed:2021}. In contrast, the CREX experiment, which analyzed the neutral weak form factor of ${ }^{48}$Ca, reports a much lower value of $L_{\text{sym}}=20 \pm 30$~MeV~\citep{CREX:2022}. A combined analysis of PREX and CREX data tends to favor an intermediate value near $L_{\text{sym}} \approx 60$~MeV~\cite{Lattimer:2023}. Additionally, measurements of charged pion spectra at high transverse momenta in heavy-ion collisions constrain $L_{\text{sym}}$ to the range $42<L_{\text{sym}}<117$~{MeV}~\cite{SpiRIT:2021}. More recently, ultrarelativistic collisions at the LHC, involving the determination of the neutron skin thickness of ${ }^{208}$Pb, have yielded a somewhat higher value of $L_{\text{sym}}=79 \pm 39$~MeV~\cite{Giacalone:2023}.

Therefore, comparing inferences based on different likelihood functions allows us to test the consistency of various statistical approaches and ultimately establish tighter constraints on the EOS across the full relevant density range. 
By examining how different statistical treatments affect conclusions, we can better understand which constraints are most susceptible to the methodology, potentially leading to more reliable determinations of nuclear matter properties from low densities to the extreme densities found in CSs.

This paper is structured as follows. In Sec.~\ref{sec:Model} we introduce the specific EOS model employed in this work. In Sec.~\ref{sec:Bayesian} we discuss the Bayesian inference framework and define the various likelihood functions adopted for various constraints. Our analysis of the impacts of different choices of the likelihood functions on the predictions of the model is given in Sec.~\ref{sec:Results}. Here we present the results of a comparison of gross properties of CSs and nucleonic matter. Our conclusions are given in Sec.~\ref{sec:Conclusions}.

%----------------------------------------------------------
\section{CDF for stellar matter}
\label{sec:Model}
%----------------------------------------------------------
To ensure this presentation is self-contained, we briefly review 
the general feature of CDF approach, also given in the first two 
papers of this series~\cite{Lijj:2025a,Lijj:2025b} and reviewed 
in Ref.~\cite{Sedrakian:2023}. We use the meson-exchange version 
with density dependent couplings. The Lagrangian of stellar matter 
with nucleonic degrees of freedom reads
$\mathscr{L} = \mathscr{L}_N + \mathscr{L}_m + \mathscr{L}_l$,
where the effective nucleonic Lagrangian is given by
% --------------------------------------------------------
\begin{eqnarray}
\label{eq:Lagrangian}
\mathscr{L}_N  = 
\sum_N\bar\psi_N \Big[\gamma^\mu
\big(i \partial_\mu - g_{\omega}\omega_\mu 
    - g_{\rho} \bm{\tau} \cdot \bm{\rho}_\mu \big)
    -\big(m_N - g_{\sigma}\sigma\big) \Big]\psi_N, \nonumber \\ 
\end{eqnarray}
% --------------------------------------------------------
where $\psi_N$ are the nucleonic Dirac fields with masses $m_N$, 
and $\sigma,\,\omega_\mu$, and $\bm{\rho}_\mu$ are the mesonic 
fields that mediate the nuclear interactions. The remaining pieces 
of the Lagrangian correspond to the mesonic and leptonic contributions, 
respectively. In practice, the masses of nucleons, mesons and leptons 
are fixed to be (or be close to) the ones in the vacuum~\cite{Lijj:2025b}. 

The meson-nucleon couplings depend on the vector nucleonic 
density $\rho = \langle \psi^\dagger \psi \rangle$,
which were assumed as
\begin{align}
g_{m}(\rho)=g_{m}(\rho_{\rm {sat}})f_m(r),
\end{align}
with a constant value $g_{m}(\rho_{\rm{sat}})$ at saturation 
density $\rho_{\rm{sat}}$, and a function $f_m(r)$ of the ratio 
$r=\rho/\rho_{\rm{sat}}$. For the isoscalar channel, the density 
dependence is defined as~\cite{Typel:1999,Lalazissis:2005}
\begin{align}\label{eq:isoscalar_coupling}
f_{m}(r)=a_m\frac{1+b_m(r+d_m)^2}{1+c_m(r+d_m)^2}, \quad 
m = \sigma,\,\omega,
\end{align}
with five conditions:
\begin{align}\label{eq:coupling_constraints}
f_{m}(1)=1, \quad f^{\prime\prime}_{m}(0)=0, \quad
f^{\prime\prime}_{\sigma}(1)=f^{\prime\prime}_{\omega}(1),
\end{align}
which reduce the number of adjustable parameters.
For the isovector channel the density dependence 
is taken in an exponential form~\cite{Typel:1999,Lalazissis:2005}:
\begin{align}\label{eq:isovector_coupling}
f_{\rho}(r) = \text{exp}\,[-a_\rho (r-1)].
\end{align}

To describe cold, neutrino-free, and catalyzed stellar matter, 
we further require weak equilibrium and charge neutrality that 
prevail in CSs~\cite{Sedrakian:2023}. The EOS of stellar matter 
and the corresponding composition, then, can be computed uniquely 
in terms of seven adjustable parameters 
\begin{align}\label{eq:cdf_parameters}
\bm{\theta}_{\rm EOS} = (g_\sigma,\,g_\omega,\,g_\rho,\,
a_\sigma,\,d_\sigma,\,d_\omega,\,a_\rho),
\end{align}
that enter our Bayesian inference. These are the three 
coupling constant at nuclear saturation density 
($g_\sigma,\,g_\omega,\,g_\rho$) and four parameters 
($a_\sigma,\,d_\sigma,\,d_\omega,\,a_\rho$) that control 
their density dependences, for those we take uniform 
prior distributions with the intervals listed in 
Table~\ref{tab:CDF_parameters}.

%----------------------------------------------------------
\begin{table}[tb]
\centering
\caption{The  minimal (``min'') and maximal
(``max'') values of the variation range of 
the CDF parameters over which uniform prior 
distributions have been assigned. 
}
\setlength{\tabcolsep}{20.4pt}
\label{tab:CDF_parameters}
\begin{tabular}{cccc}
\hline\hline
No.& Parameter  & min & max  \\
\hline
1  & $g_\sigma$ & 7.5 & 12.5 \\
2  & $g_\omega$ & 9.0 & 16.0 \\
3  & $g_\rho$   & 2.5 &  5.0 \\
4  & $a_\sigma$ & 0.8 &  2.2 \\
5  & $d_\sigma$ & 0.0 &  2.2 \\
6  & $d_\omega$ & 0.0 &  2.2 \\
7  & $a_\rho$   & 0.0 &  1.5 \\
\hline\hline
\end{tabular}
\end{table}
%----------------------------------------------------------

%----------------------------------------------------------
\section{Inference framework and constraints}
\label{sec:Bayesian}
%----------------------------------------------------------
We perform a Bayesian analysis for the CDF approach by applying nuclear 
and astrophysical constraints. The details of the techniques used can 
be found in Refs.~\citep{Lijj:2025a,Lijj:2025b}. For the sake of 
completeness, here we present a few key aspects and highlight 
additional changes made for the likelihood functions.

%----------------------------------------------------------
\subsection{Low density matter constraints and the likelihoods}
%----------------------------------------------------------
The bulk properties of nuclear matter are customarily quantified 
in terms of the {\it nuclear matter characteristics at saturation density}, 
which are the coefficients of the Taylor expansion of the energy density 
close to the saturation density and isospin-symmetric limits
\begin{eqnarray}
\label{eq:Taylor_expansion}
\varepsilon\,(\rho, \delta) 
&\simeq&
E_{\rm{sat}} + \frac{1}{2!}K_{\rm{sat}}\chi^2
+ \frac{1}{3!}Q_{\rm{sat}}\chi^3 
+ \frac{1}{4!}Z_{\rm{sat}}\chi^4 \nonumber \\ [1.0ex]
& & +\,J_{\rm{sym}}\delta^2 + L_{\rm{sym}}\delta^2\chi
+ \frac{1}{2!}K_{\rm{sym}}\delta^2\chi^2 
+ {\mathcal O}(\chi^5,\chi^3\delta^2), \nonumber \\
\end{eqnarray}
where $\rho=\rho_n + \rho_p$ is the baryonic density 
with $\rho_{n(p)}$ denoting the neutron (proton) density, 
$\chi=(\rho-\rho_{\rm{sat}})/3\rho_{\rm{sat}}$, and 
$\delta=(\rho_{n}-\rho_{p})/\rho$ is the isospin asymmetry.

The coefficients in the first line of the 
expansion~\eqref{eq:Taylor_expansion} are known as 
isoscalar characteristics, namely, the {\it saturation energy} 
$E_{\rm{sat}}$, the {\it incompressibility} $K_{\rm{sat}}$, 
the {\it skewness} $Q_{\rm{sat}}$, and the {\it kurtosis} 
$Z_{\rm{sat}}$. The isovector characteristics (in the second line) 
associated with the density dependence of symmetry energy are 
the {\it symmetry energy parameter} $J_{\rm{sym}}$, its 
{\it slope parameter} $L_{\rm{sym}}$, and the {\it curvature} 
$K_{\rm{sym}}$. The higher-order terms in the expansion~\eqref{eq:Taylor_expansion}, 
which are not shown here, have been studied, for example, in 
Refs.~\citep{Margueron:2018a,Margueron:2019,Providencia:2023,Char:2023}.

To further specify the properties of a CDF, one additionally needs 
the value of the Dirac effective mass $M^\ast_{\rm D}$ at the 
saturation density, which is important for a quantitative 
description of the physics of finite nuclei~\citep{Lijj:2014}. 
Consequently, in the present CDF framework, the seven adjustable 
parameters in Eq.~\eqref{eq:cdf_parameters} can be converted to 
seven macroscopic characteristics for symmetric nuclear matter (SNM),
\begin{align}\label{eq:eos_parameters}
\bm{\theta}_{\rm{SNM}} = (M^\ast_{\rm{D}},\,\rho_{\rm{sat}},\,
E_{\rm{sat}},\,K_{\rm{sat}},\,Q_{\rm{sat}},\,J_{\rm{sym}},\,L_{\rm{sym}}).
\end{align}
In particular, one could individually consider one of the quantities 
in Eq.~\eqref{eq:eos_parameters} while fixing the 
others~\citep{Lijj:2019a,Lijj:2019b,Lijj:2023b}. The mean 
values and standard deviations, or the respective parameter intervals, 
suggested by various works are summarized in Table~\ref{tab:NM_constraints}.
To be more specific, the mean values and standard
deviations for those lower-order parameters are extracted from
statistical analysis of distributions provided by collections 
of nuclear models~\citep{Dutra:2014,Margueron:2018a}.
Notably, we consider a broad range for $Q_{\text{sat}}$ and 
$L_{\text{sym}}$ that reflects the uncertainties of their values 
that exist in the literature~\cite{PREX:2021,CREX:2022,Reed:2021,Reinhard:2021,Dutra:2014,Margueron:2018a,Lijj:2023b,Sunboyang:2024}.
However, it should be noted that not all combinations of parameter 
values (within the prior ranges) predict good results for finite 
nuclei given a specific nuclear model.

%----------------------------------------------------------
\begin{table}[tb]
\centering
\caption{
Symmetric nuclear matter (SNM) characteristics at saturation 
density and pure neutron matter (PNM) properties from $\chi$EFT 
used to constrain the CDF parameters.} 
\setlength{\tabcolsep}{9.2pt}
\label{tab:NM_constraints}
\begin{tabular}{ccccccc}
\hline\hline
     & No.& Quantity         & Unit        & Interval           \\
\hline
SNM  & 1  &$M_{\rm D}^\ast$  & $m_{\rm N}$ & $ 0.60 \pm 0.05$  \\
     & 2  &$\rho_{\rm sat}$  & fm$^{-3}$   & $ 0.153 \pm 0.005$\\
     & 3  &$E_{\rm sat}$     & MeV         & $-16.1 \pm 0.2$   \\
     & 4  &$K_{\rm sat}$     & MeV         & $ 230 \pm 40$     \\
     & 6  &$J_{\rm sym}$     & MeV         & $ 32.5 \pm 2.0$   \\
     & 5  &$Q_{\rm sat}$     & MeV         & $[-1400, 1400]$   \\
     & 7  &$L_{\rm sym}$     & MeV         & $[    0,  120]$   \\
\hline
PNM  & 8  &$P\,(\rho)$       &MeV fm$^{-3}$& N$^3$LO           \\
     & 9  &$\varepsilon\,(\rho)$&MeV fm$^{-3}$& N$^3$LO        \\
\hline\hline
\end{tabular}
\end{table}
%----------------------------------------------------------

In addition, we incorporate $\chi$EFT results for pure neutron matter 
(PNM) in the density range 0.5 -- $1.1\,\rho_{\rm{sat}}$ which constrain 
further the low-density regime of nucleonic EOS. In practice, we use the 
N$^3$LO computations for the energy per particle and pressure
from Ref.~\citep{Hebeler:2013} at discrete density points 
(i.e., 0.08, 0.12 and 0.16~$\rm{fm}^{-3}$) and assume the 
uncertainty as a Gaussian distribution at $1\,\sigma$. 
Note that the uncertainty band reported in Ref.~\citep{Hebeler:2013} 
lies fully within the bands predicted by various $\chi$EFT based 
computations as compiled in Refs.~\citep{Huth:2021,Huth:2022}.
In addition, though no correlation was assumed to exist among the 
various SNM constraints, the $\chi$EFT constraint leads to a strong 
$J_{\rm sym}$-$L_{\rm sym}$ correlation~\citep{Malik:2022a,Scurto:2024,Lijj:2025a,Lijj:2025b}.

Next, we define likelihood functions for the aforementioned 
constraints for our analysis. For quantities with greater 
uncertainties, namely, $Q_{\rm{sat}}$ and $L_{\rm{sym}}$, 
we adopt a pass-band selection which rejects models featuring 
values outside that range given in Table~\ref{tab:NM_constraints}. 
For quantities that are known to within approximately $10\%$ or 
are given via the $\chi$EFT computations, we assume two types of 
prior distributions. 

(I). Gaussian distribution (abbreviated as ``Gaus."). 
We assume a Gaussian prior distribution for each quantity, 
and the likelihood function reads
%
%--------------------------------
\begin{align}\label{eq:Gaussian function}
\mathcal{L}_{\rm NM}\,(\bm\theta_{\rm EOS}) = 
\frac{1}{\sqrt{2\pi}\,\sigma}
{\rm exp} \left[-\frac{1}{2}
\left(\frac {\mu(\bm \theta_{\rm EOS})-d}{\sigma}\right)^2\right],
\end{align}
%--------------------------------
%
where $d$ and $\sigma$ are the datum and its standard uncertainty, 
respectively, and $\mu$ the corresponding model value.

(II). Uniform-Gaussian combination (abbreviated as UG). 
We assume a flat distribution that is centered on the assigned 
mean value for each quantity with a width of $2\,\sigma$ or 
$4\,\sigma$ interval, where the $1\,\sigma$ value is the one 
listed in Table~\ref{tab:NM_constraints}. Above or below the interval, 
the probabilities decay slowly as for a Gaussian function. 
Specifically, the likelihood function is defined as
%
%--------------------------------
\begin{align}\label{eq:UG function}
\mathcal{L}_{\rm NM}\,(\bm\theta_{\rm EOS}) =
\left\{  
\begin{array}{lc}  
\mathcal{L}_{\rm U}(\mu), & d_{\rm min} \leq \mu \leq d_{\rm max} \\  
\mathcal{L}_{\rm G}(\mu), & {\rm otherwise} 
\end{array},
\right.  
\end{align}
%--------------------------------
%
where $\mathcal{L}_{\rm G}(\mu)$ is the Gaussian 
function defined in Eq.~\eqref{eq:Gaussian function}, and 
$\mathcal{L}_{\rm U}(\mu)$ is a constant for which we 
introduce two cases. In UG1, 
%
%--------------------------------
\begin{align}
\mathcal{L}_{\rm U}(\mu) = \frac{0.6827}{2\,\sigma},\quad
d-\sigma \leq \mu \leq d +\sigma,
\end{align}
%--------------------------------
%
and for UG2,
%
%--------------------------------
\begin{align}
\mathcal{L}_{\rm U}(\mu) = \frac{0.9545}{4\,\sigma},\quad
d-2\sigma \leq \mu \leq d+2\sigma.
\end{align}
%--------------------------------
%
In this way, we do not immediately discard all the CDF models that 
are outside the $1\,\sigma$ or $2\,\sigma$ limit preassigned for the 
SNM quantities in Table~\ref{tab:NM_constraints}, and the samples are 
spanned in the same parameter space as the case with a Gaussian function. 
Furthermore, the distribution function~\eqref{eq:UG function} remains 
normalized, which thus allows the implementation of a quantitative 
comparison of the results obtained from different likelihoods.
Finally, we stress that for each parameter, the values and/or intervals 
listed in Table~\ref{tab:NM_constraints} are used only as references to 
construct the likelihood functions. Their posterior distributions are 
updated iteratively within a Bayesian analysis process.

%----------------------------------------------------------
\subsection{Astrophysical constraints and the likelihoods}
%----------------------------------------------------------
%----------------------------------------------------------
\begin{table*}[tb]
\centering
\caption{
The astrophysical constraints used for the scenarios in the 
present analysis. For the four NICER sources, the pulse profile 
models used to generate the samples are indicated.
}
\setlength{\tabcolsep}{10.4pt}
\label{tab:Scenarios}
\centering
\begin{tabular}{ccccccccccccccc}
\hline\hline
\multirow{2}*{Scenario}  &\multirow{2}*{J0348} & GW & GW & J0740 & J0437 & 
\multicolumn{2}{c}{J0030}&\multicolumn{2}{c}{J1231}  \\
\cline{7-8} \cline{9-10}
&       & 170817 & 190425 & ST-U & CST+PDT & ST+PDT & PDT-U & PDT-U (i)& PDT-U (ii)       \\
\hline\hline
Baseline&$\times$&$\times$&        &        &        &        &        &        &         \\
B       &$\times$&$\times$&$\times$&$\times$&$\times$&$\times$&        &$\times$&         \\
F       &$\times$&$\times$&$\times$&$\times$&$\times$&        &$\times$&        &$\times$ \\
\hline
\hline
\end{tabular}
\end{table*}
%----------------------------------------------------------

We next describe our implementations of the likelihoods for various 
CS measurements, and assume that the datasets of different observed 
sources are independent. 

\underline{Massive pulsars data (MP).}
The mass measurements of massive pulsars establish a rigorous 
lower bound on the maximum mass of a static CS, $M_{\rm max}$. 
We choose the most accurate mass value measured through 
Shapiro delay for PSR J0348+0432~\citep{Antoniadis:2013},
and assume the mass as a Gaussian distribution.
The likelihood is constructed using the cumulative 
density function of the Gaussian function
%--------------------------------
\begin{align}
\mathcal{L}_{\rm MP}\,(\bm\theta_{\rm{EOS}}) = \frac{1}{2}
\left[1 + {\rm erf} \left(\frac{M_{\rm{max}}(\bm\theta_{\rm{EOS}})-M}
{\sqrt{2}\,\sigma}\right)\right],
\end{align}
%--------------------------------
where ${\rm erf}\,(x)$ is the error function, and $M$ and $\sigma$ 
are the mean and the standard deviation of the mass measurement 
for the source, respectively.

\underline{Gravitational waves data (GW).}
The GW170817~\citep{LVScientific:2017,LVScientific:2019} and 
GW190425~\citep{LVScientific:2020a} events are the only two 
confirmed—and most likely—binary neutron star mergers observed 
during previous LIGO-Virgo-KAGRA observing runs. We calculate the 
likelihood for each event through the high-precision interpolation 
in TOAST~\citep{Hernandez:2020} obtained from fitting the strain data, 
%--------------------------------
\begin{align}
\mathcal{L}_{\rm GW}\,(\bm\theta_{\rm{EOS}}) = 
F(\mathcal{M},q,\Lambda_1,\Lambda_2),
\end{align}
%--------------------------------
where $\mathcal{M} = (M_1 M_2)^{3/5}/(M_1+M_2)^{1/5}$ is the chirp
mass, $q = M_1/M_2$ is the mass ratio, and $\Lambda_1(M_1)$ and
$\Lambda_2(M_2)$ the tidal deformabilities of the individual 
stars, respectively.

\underline{NICER data.}
The NICER collaboration has provided simultaneous measurements of 
mass and radius for four millisecond pulsars through pulse profile 
modeling: a massive $\sim 2.1\,M_{\odot}$ star PSR J0740+6620~\citep{Salmi:2024a}, two 
canonical-mass $\sim 1.4\,M_{\odot}$ stars PSR J0030+0451~\citep{Vinciguerra:2024} and 
J0437-4715~\citep{Choudhury:2024}, and a low-mass $\sim 1.1\,M_{\odot}$ 
one PSR J1231-1411~\citep{Salmi:2024b}. 
We construct our likelihood function for each of the sources 
using the Gaussian kernel density estimation (KDE) with the 
released model posterior samples $\bm S$,
%--------------------------------
\begin{align}
\mathcal{L}_{\rm NICER}\,(\bm\theta_{\rm{EOS}})= {\rm KDE}(M,R|\bm{S}),
\end{align}
%--------------------------------
where the mass $M$ and radius $R$ for the star are functions of 
its central pressure (equivalently, central density) and of the sampled EOS parameters.

In the present analysis, we implement the reported $(M,\,R)$ samples 
as listed in Table~\ref{tab:Scenarios}. For details of the pulse profile 
modelings, see Refs.~\citep{Choudhury:2024,Vinciguerra:2024,Salmi:2024a,Salmi:2024b}.
Note that the reanalysis of PSR J0030+0451 resulted in three 
different ellipses in the \MR plane, each corresponding to a 
different pulse profile modeling~\citep{Vinciguerra:2024}. 
For PSR J1231-1411, the inference results show a strong sensitivity to 
the choice of radius priors, with stable and likely converged outcomes achieved only under constrained radius 
priors~\citep{Salmi:2024b}. In this context, we propose scenarios B 
and F in Table~\ref{tab:Scenarios} to represent two distinct combinations of the current NICER estimates for pulsars, corresponding to the softest and stiffest models, respectively (see also Refs.~\citep{Lijj:2025a,Lijj:2025b}). Additionally, we introduce the Baseline scenario, which incorporates a minimal set of astrophysical constraints—specifically, the mass of PSR J0348+0432 and the tidal deformabilities from GW170817. Below, we perform illustrative tests 
using these three scenarios.

%----------------------------------------------------------
\section{Results and implications}
\label{sec:Results}
%----------------------------------------------------------
We now assess the current theoretical uncertainties arising from the choice of likelihood functions used to constrain low-density matter and examine their implications for the properties of CSs and dense matter. Our statistical analysis is based on approximately $3 \times 10^4$ posterior EOS models that satisfy both nuclear theory and observational constraints from pulsar and gravitational-wave measurements.

%----------------------------------------------------------
\subsection{Properties of compact stars}
%----------------------------------------------------------

%--------------------- MR S-mini ------------------
\begin{figure}[b]
\centering
\includegraphics[width = 0.44\textwidth]{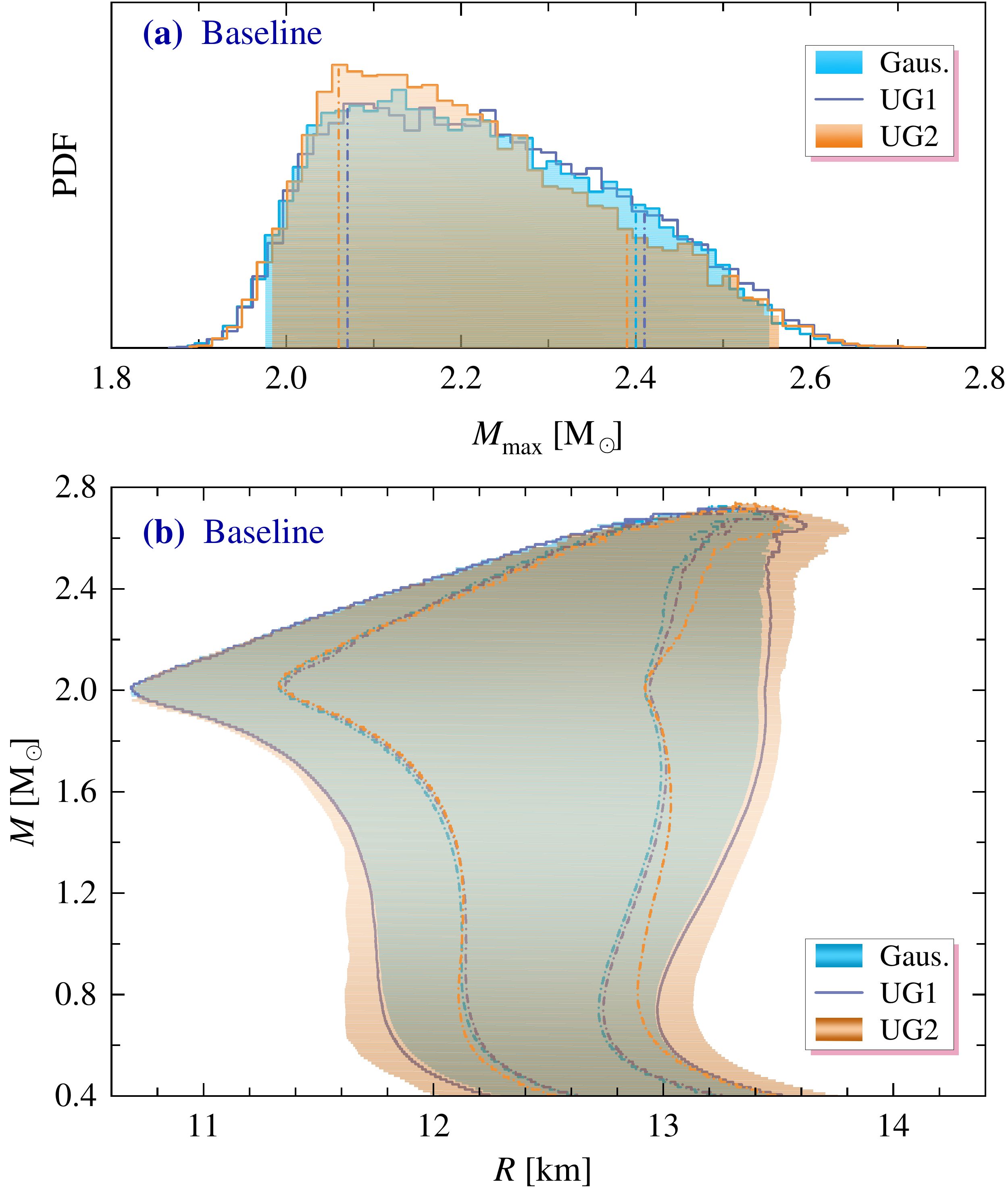}
\caption{
Posterior distributions for mass-radius relation and maximum mass under the Baseline astrophysical scenario, assuming low-density matter constraints are modeled using either a Gaussian likelihood (Gaus.) or a hybrid of uniform and Gaussian distributions (UG1 and UG2). Shaded regions indicate the 95.4\% CIs, while solid lines represent the 68.3\% CIs.
}
\label{fig:MR_Baseline}
\end{figure}
%--------------------- MR S-mini ------------------

%-------------------- MR S-BF ---------------------
\begin{figure*}[tb]
\centering
\includegraphics[width = 0.96\textwidth]{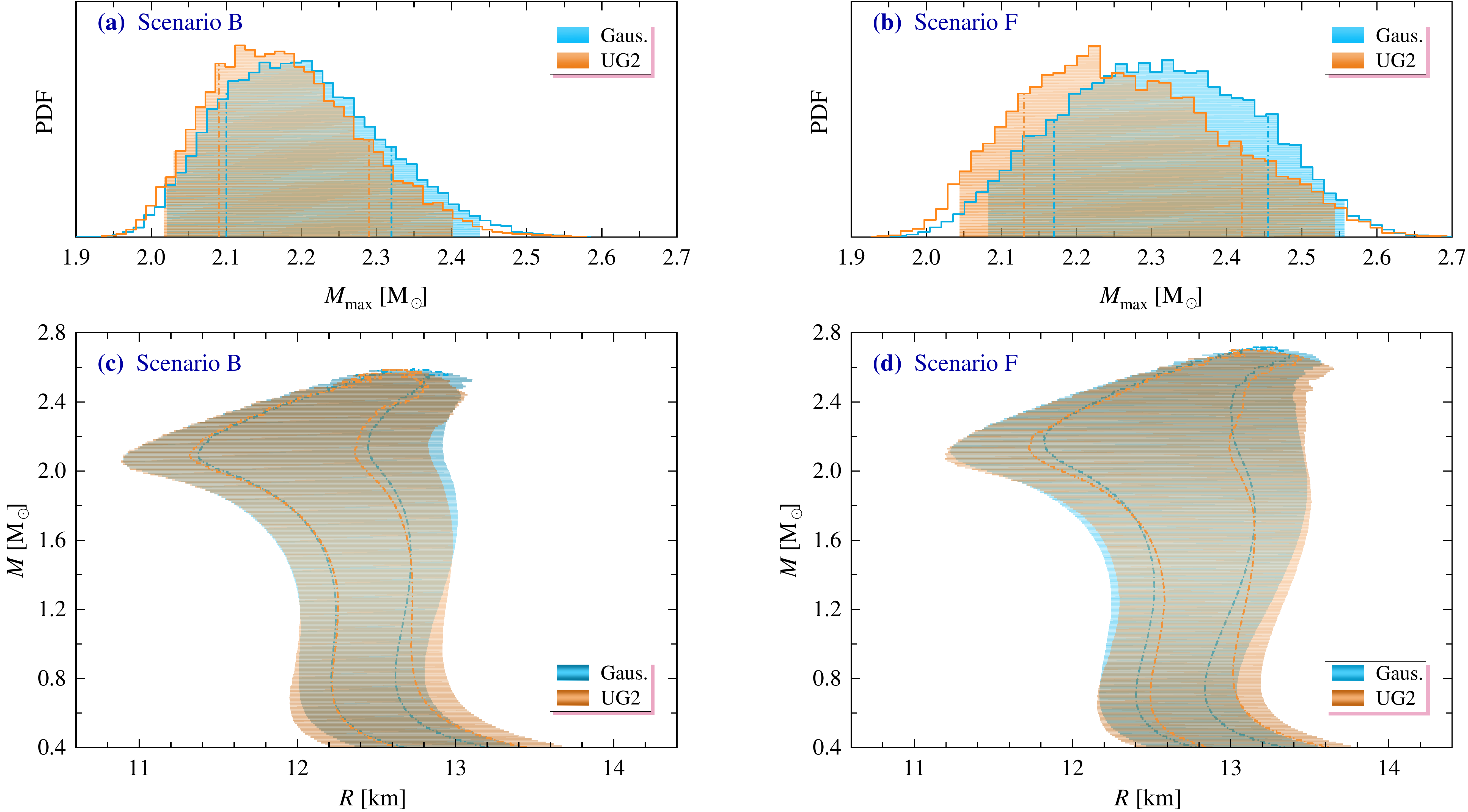}
\caption{
Posterior distributions for the mass-radius relation and maximum 
mass under astrophysical scenarios B (left panels) and F (right panels), assuming low-density matter constraints modeled using 
either a Gaussian likelihood (Gaus.) or a combined uniform-Gaussian distribution (UG2). Shaded regions correspond to 95.4\% CIs, while the lines indicate 68.3\% CIs.
}
\label{fig:MR_BF}
\end{figure*}
%-------------------- MR S-BF ---------------------

We begin by evaluating the impacts of different likelihood choices for low-density matter constraints on the CS properties, assuming either a Gaussian prior (Gaus.) or a combination of uniform and Gaussian priors (UG1 or UG2). The bulk properties of CSs — such as their maximum mass, radii, tidal deformabilities, and moments of inertia — are entirely determined by the distributions of energy density and pressure throughout the star, as prescribed by the EOS of dense matter. However, these properties are integrated quantities, meaning they are sensitive to how pressure and energy density are distributed across the full range of densities within the star.
 
In Fig.~\ref{fig:MR_Baseline}, we present the \MR posteriors
and the probability distribution functions (PDFs) for maximum mass $M_{\rm max}$ under scenario ``Baseline", for which the 68.3\% and 95.4\% confidence intervals (CIs) are indicated. 
Physically, the mass $\sim 2\,M_{\odot}$ of PSR J0348+0432 sets 
a lower bound on the maximum mass $M_{\rm max}$, and thus provides the most rigorous constraint on the high-density behavior of the EOS of dense stellar matter; the tidal deformabilities deduced from GW170817 event places additional constraints on the intermediate range of EOS. It is clearly seen from Fig.~\ref{fig:MR_Baseline} that the two posteriors are almost identical for computations with Gaussian and UG1 likelihood functions, both at 68.3\% and 95.4\% CIs. Such consistency is further established from the specific values of key gross quantities of CSs and the characteristic parameters of nuclear matter at saturation density, which are collected in the Appendix. The differences for each quantity between the 
two approaches are well within 1\%. A similar conclusion was reached in Ref.~\citep{Scurto:2024}, where the two types of likelihood functions were applied exclusively to the $\chi$EFT constraints.

In light of the above results, from now on, our comparative study 
shall be concentrated on the computations with a wider 
uniform distribution, i.e., the results obtained from the UG2 
likelihood function. In this case, as shown in Fig.~\ref{fig:MR_Baseline}, the \MR posterior region broadens noticeably by approximately 0.2 km on both sides. For masses $M \lesssim 1.6\,M_{\odot}$, this widening is primarily due to the broader range of EOSs at low densities ($\rho \lesssim 2\,\rho_{\rm ast}$) permitted by the looser $\chi$EFT constraint under the UG2 likelihood. This allows models with more extreme values of the isovector parameters $J_{\rm sym}$ and $L_{\rm sym}$ to contribute to the posterior — parameters that play a key role in determining the radii of low-mass stars.

%--------------------- EOS cs ---------------------
\begin{figure*}[tb]
\centering
\includegraphics[width = 0.96\linewidth]{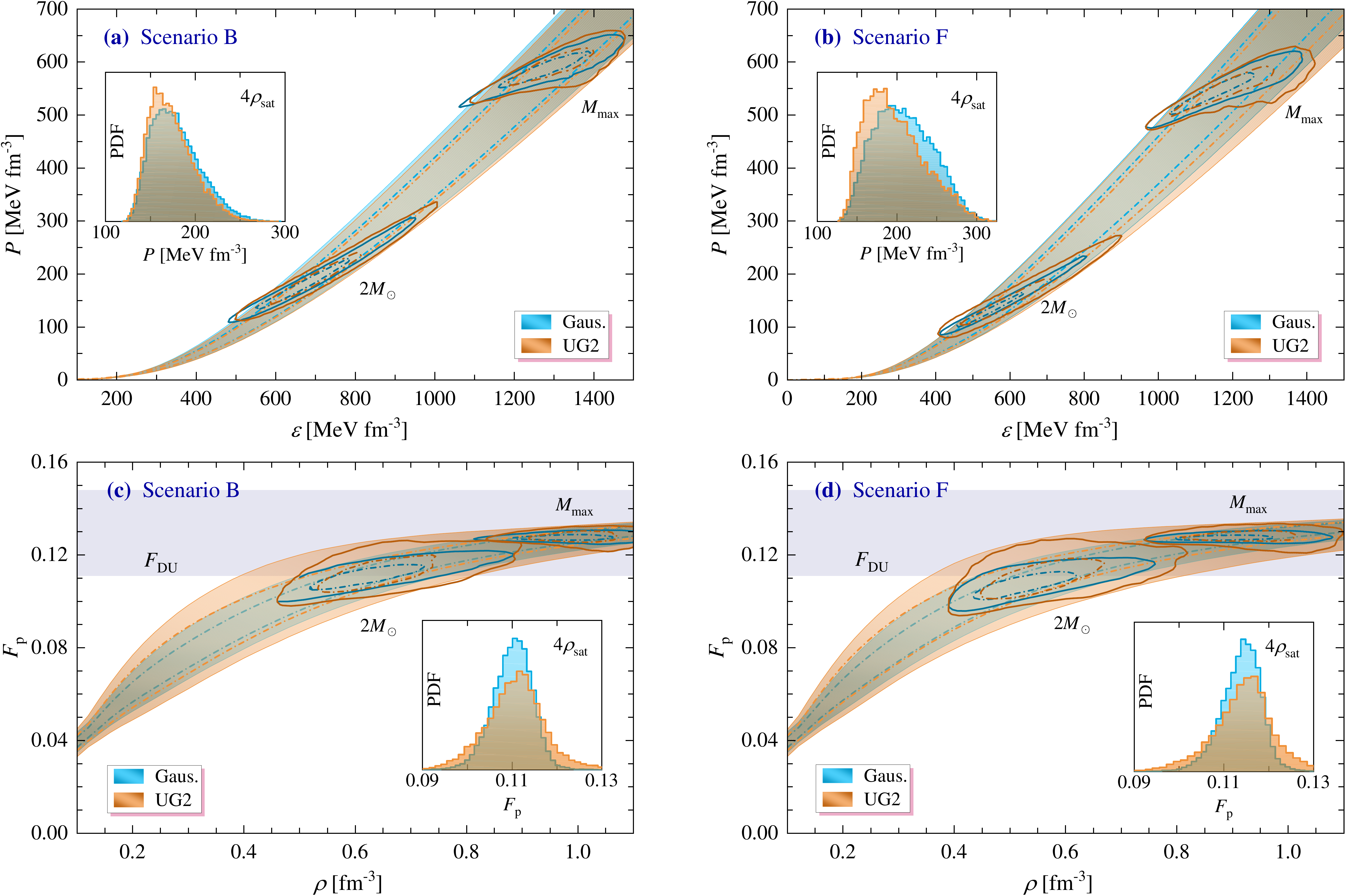}
\caption{
Posteriors for $\beta$-stable equation of state and proton 
fraction distributions under astrophysical scenarios B 
(left panels) and F (right panels), assuming for the 
low-density matter constraints a likelihood of Gaussian 
distribution (Gaus.) or a combined uniform-Gaussian 
distribution (UG2). The shaded regions represent the 
distributions at 95.4\% CI while the lines for 68.3\% CI. 
In each panel, the contours indicate the corresponding 
distributions of the respective $2.0\,M_{\odot}$ and 
the maximum mass $M_{\rm max}$ configurations, and the 
insets show the probability density functions (PDFs) of 
the pressure or fraction at $4\,\rho_{\rm sat}$. In lower 
panels the orange bands labeled $F_{\rm DU}$ show the 
admissible threshold values for the onset of nucleonic 
direct Urca (DU) cooling process.
}
\label{fig:EOS_cs}
\end{figure*}
%--------------------- EOS cs ---------------------

%--------------------- EOS nm ---------------------
\begin{figure*}[tb]
\centering
\includegraphics[width = 0.96\linewidth]{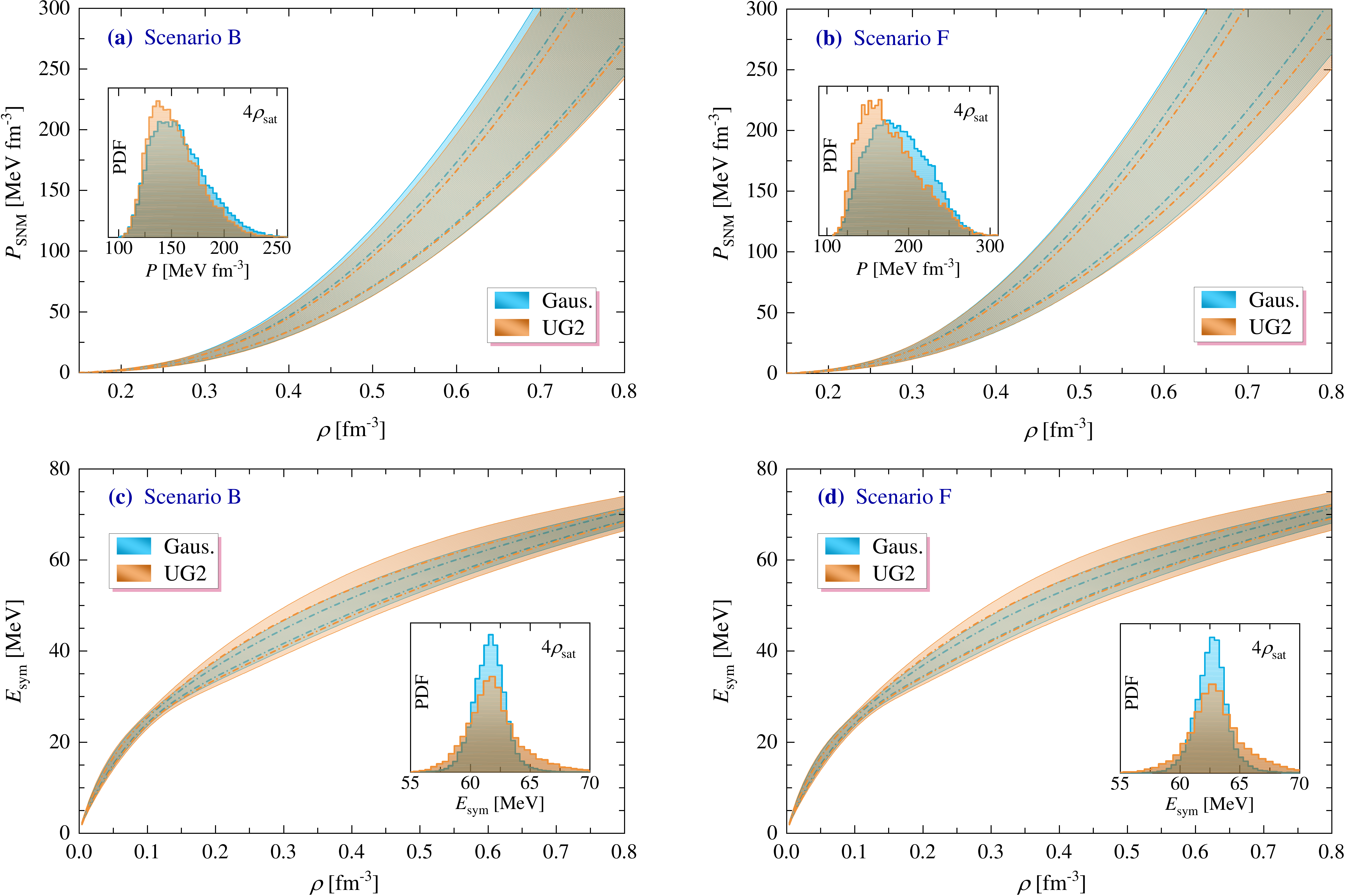}
\caption{
Posterior distributions for nucleonic equation of state and symmetry energy under astrophysical scenarios B (left panels) 
and F (right panels), assuming low-density matter constraints are modeled using either a Gaussian likelihood (Gaus.) or a combined uniform-Gaussian distribution (UG2). Shaded regions indicate the 95.4\% CI, while lines represent the 68.3\% CI. Insets in each panel display the probability density functions (PDFs) of the pressure or symmetry energy at $4\,\rho_{\rm sat}$.
}
\label{fig:EOS_nm}
\end{figure*}
%--------------------- EOS nm ---------------------

%----------------- NM characteristics -------------
\begin{figure*}[tb]
\centering
\includegraphics[width = 0.96\textwidth]{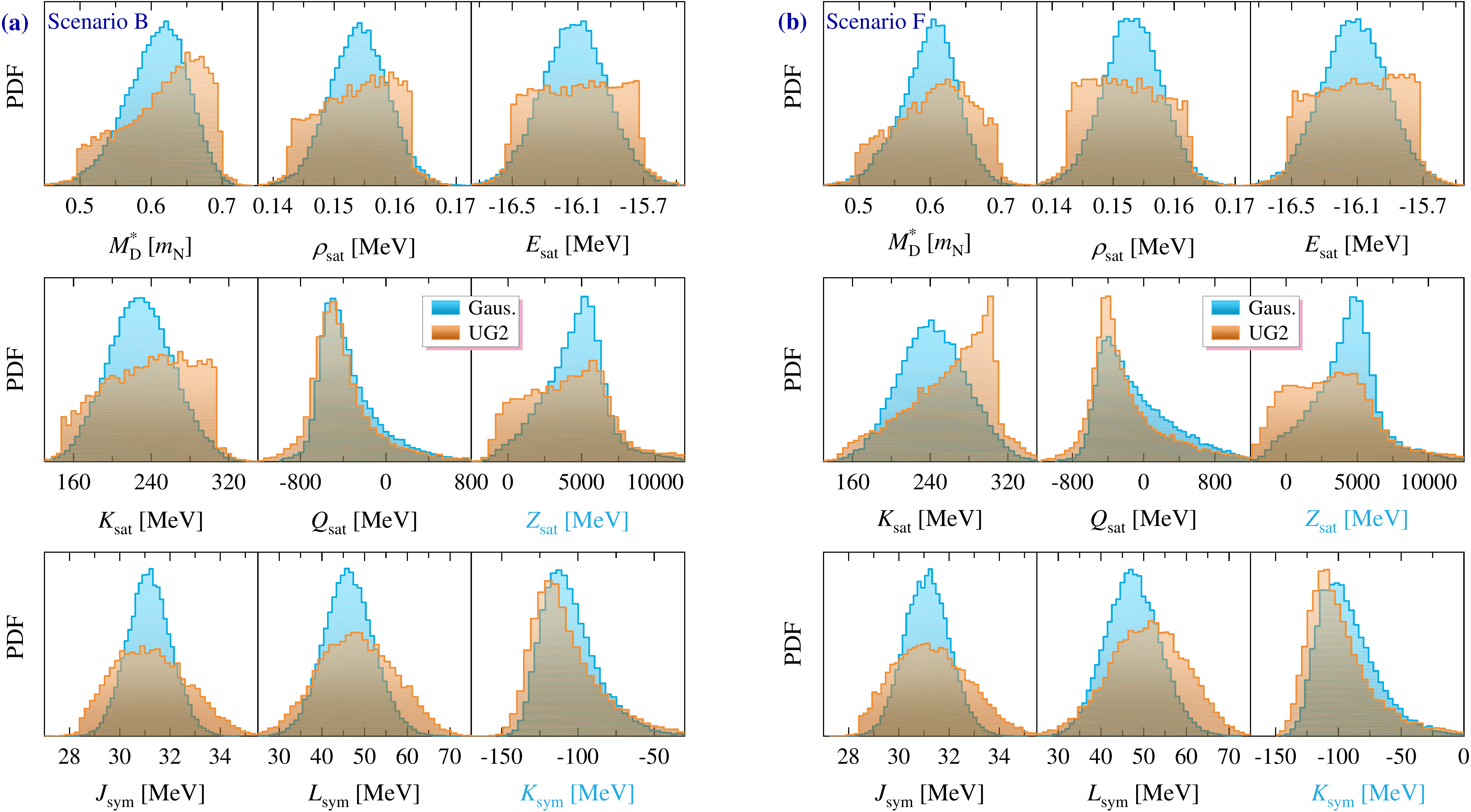}
\caption{Posterior distributions for nuclear matter properties at saturation density under astrophysical scenarios B (left panels) and F (right panels), assuming low-density matter constraints modeled using either a Gaussian likelihood (Gaus.) or a combined uniform-Gaussian distribution (UG2). The higher-order parameters $Z_{\rm sat}$ and $K_{\rm sym}$, shown in blue, are predicted from the 
lower-order parameters within the present CDF framework.
}
\label{fig:NM_charact}
\end{figure*}
%----------------- NM characteristics -------------

Meanwhile, a lower value of $L_{\rm sym}$ can be offset by a higher value of the isoscalar skewness parameter $Q_{\rm sat}$~\citep{Lijj:2019a}, resulting in a viable model within the current framework. The parameter $Q_{\rm sat}$ predominantly influences the radii of high-mass stars ($M \gtrsim 1.6\,M_{\odot}$) as well as the maximum mass of CSs~\citep{Lijj:2019b}. The posterior distributions of the nuclear matter characteristics will be discussed in the following subsection.

The combined NICER and XMM-Newton measurements of both the radii and masses of four pulsars~\citep{Riley:2019,Riley:2021,Miller:2019,Miller:2021,Choudhury:2024,Vinciguerra:2024,Salmi:2024a,Salmi:2024b}, with masses ranging from approximately $1.0$ to $2.1\,M_{\odot}$, along with the tidal deformabilities inferred from the GW170817 and GW190425 binary neutron star merger events~\citep{LVScientific:2017,LVScientific:2020a}, provide a valuable opportunity to constrain the dense matter EOS at densities between $2$ and $4\,\rho_{\rm sat}$. We present illustrative results for scenarios B and F in Fig.~\ref{fig:MR_BF}, after incorporating NICER data for four pulsars into our analysis. In scenario B, we use more compact estimates for PSR J0030+0451 (ST+PDT model) and J1231-1411 (PDT-U (i) model), which favor softer EOSs at densities below $\sim 2$ and $3\,\rho_{\rm sat}$, respectively, resulting in the tightest credible regions. In contrast, scenario F, which incorporates less compact estimates for PSR J0030+0451 (PDT-U model) and J1231-1411 (PDT-U (ii) model), predicts the widest credible regions. In the lower panels of Fig.~\ref{fig:MR_BF}, we observe that the impacts of the likelihood functions for low-density matter constraints on the \MR distributions are largely suppressed. With a Gaussian prior (Gaus.) or a combination of uniform and Gaussian distributions (UG2), the differences in radii for stars with masses $M \gtrsim 1\,M_{\odot}$ are well within 0.1 km. In contrast, for computations using UG2 likelihood functions, the predicted maximum masses $M_{\rm max}$ are shifted to lower values by at most $0.05\,M_{\odot}$ compared to those obtained using Gaussian likelihoods, as shown in the upper panels of Fig.~\ref{fig:MR_BF}. Additionally, in scenario F, computations with UG2 likelihood functions still permit maximum masses exceeding $2.5\,M_{\odot}$ at a 95.4\% CI, thus supporting a static CS interpretation for the secondary component of the GW190814 event~\citep{LVScientific:2020b}.

The inferred key quantities of CSs, such as radii, tidal deformabities, central baryonic densities, energy densities, pressures, sound speeds $(c^2_s = d P/d \varepsilon)$, and the dimensionless trace anomalies $(\Delta = 1/3-P/\varepsilon)$~\citep{Fujimoto:2022} for $1.0,\,1.4,\,2.0\,M_{\odot}$ and the maximum-mass configurations are collected in the Appendix.

Figure~\ref{fig:EOS_cs} summarizes the posterior distributions for $\beta$-equilibrium EOS that generated the posteriors 
in Fig.~\ref{fig:MR_BF}, as well as the corresponding proton fraction. In each panel, we also show the contours representing the positions of the respective $2.0\,M_{\odot}$ and the maximum-mass $M_{\rm max}$ configurations, and the PDFs of 
the pressure or fraction at a density $4\,\rho_{\rm sat}$. In the lower panels of Fig.~\ref{fig:EOS_cs} the horizontal bands indicate
the nucleonic direct Urca (DU) threshold for the CS cooling process. The lower limit of 11.1\% is derived from the $\mu^-$ free model, while the upper limit comes from a model which assumes massless $\mu$, which, however, applies only in high-density matter~\citep{Klahn:2006}.
 
As seen in the upper panels of Fig.~\ref{fig:EOS_cs}, both computations for scenarios B and F yield consistent results with Gaussian or UG2 likelihood functions. The analyses with UG2 likelihoods favor a slightly softer EOS, as indicated by the insets, where the peaks of the pressure PDFs are shifted toward lower values for a fixed baryonic density.

As shown in the lower panels of Fig.~\ref{fig:EOS_cs}, the analyses with UG2 likelihoods result in a broader proton fraction distribution, due to modifications in the value of the symmetry energy slope $L_{\rm sym}$ (which will be discussed in the next section). However, these modifications do not alter our previous conclusion~\citep{Lijj:2025a,Lijj:2025b} that the nucleonic DU process is largely disallowed in CSs with $M < 2.0\,M_{\odot}$ when using the current CDF-based EOS. A similar conclusion was reached in Refs.~\citep{Malik:2022a,Char:2025}, where the density dependence of the $\rho$-meson coupling was also modeled using an exponential form, see Eq.~\eqref{eq:isovector_coupling}.

%----------------------------------------------------------
\subsection{Properties of nucleonic matter}
%----------------------------------------------------------
%---------------------- Heatmap -------------------
\begin{figure*}[tb]
\centering
\includegraphics[width = 0.96\textwidth]{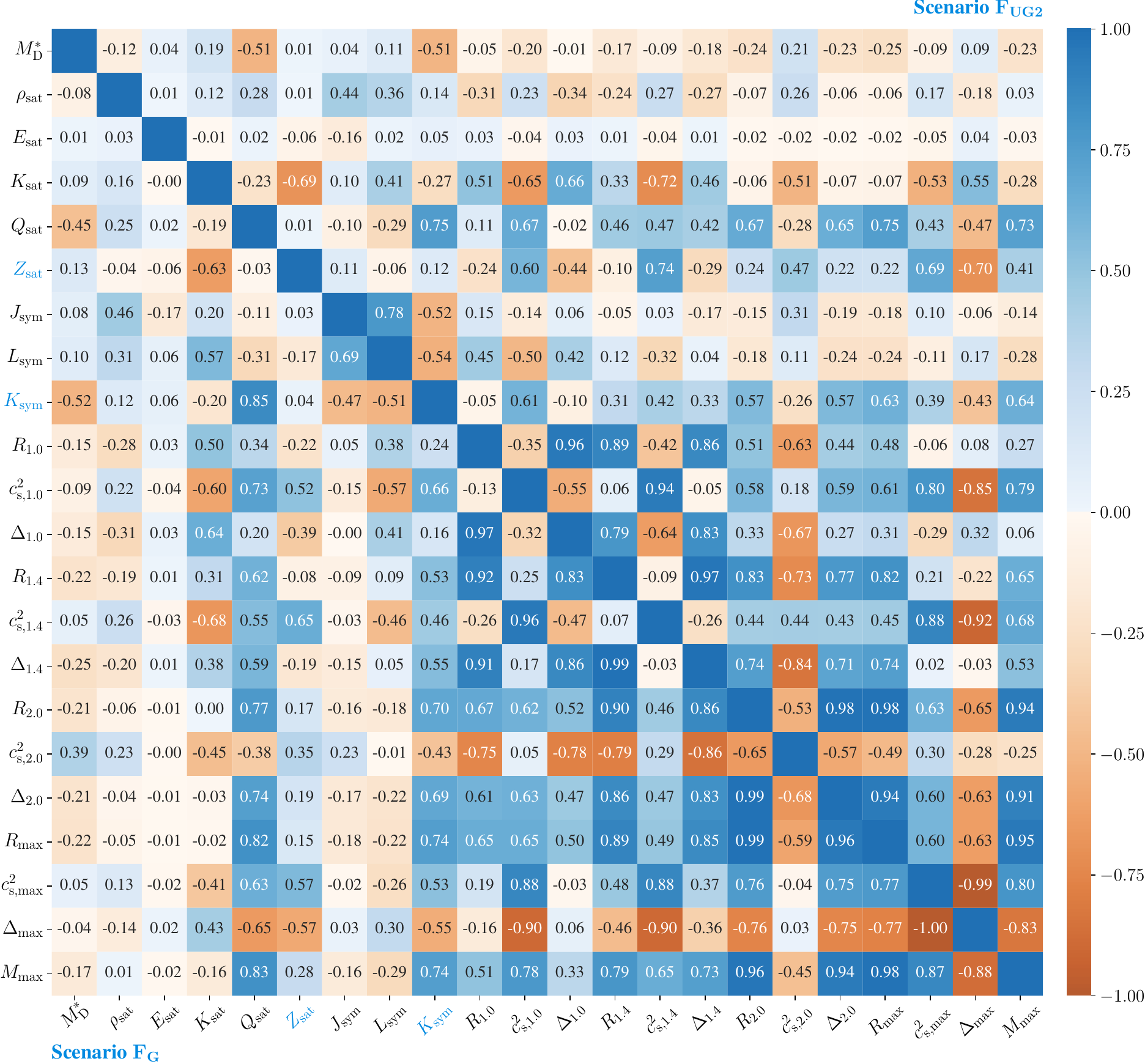}
\caption{
Posterior correlation matrix showing the variation of nuclear 
characteristic parameters at saturation density and selected 
bulk properties of compact stars under astrophysical scenario 
F. The lower-left part corresponds to constraints modeled with 
a Gaussian likelihood (G), while the upper-right reflects a 
combined uniform-Gaussian distribution (UG2). Higher-order 
parameters $Z_{\rm sat}$ and $K_{\rm sym}$, highlighted in 
blue, are predicted from lower-order parameters within the 
current CDF framework.
}
\label{fig:Heatmap}
\end{figure*}
%---------------------- Heatmap -------------------

Next, we assess the impacts of the likelihoods for low-density matter 
constraints on the properties of symmetric nuclear matter and the 
density dependence of symmetry energy.

In Fig.~\ref{fig:EOS_nm}, we display the posterior distributions 
for pressure (upper panels) and symmetry energy (lower panels) as 
a function of baryonic density under astrophysical scenarios B and F. 
We assume a Gaussian distribution (Gaus.) for the low-density matter 
constraints or a combination of uniform and Gaussian distributions (UG2). 
In each panel we also illustrate the probability density functions (PDFs) 
at a density of $4\,\rho_{\rm sat}$. The upper panels of Fig.~\ref{fig:EOS_nm} 
show consistency with the $\beta$-stable EOS presented in Fig.~\ref{fig:EOS_cs}, 
indicating that analyses using the UG2 likelihood slightly favor a softer 
symmetrical matter EOS. For the symmetry energy, the UG2 likelihood 
function results in a broad distribution, particularly in high-density 
regions. However, the variations in symmetry energy values remain below 
3~MeV across the relevant density ranges, which may not significantly impact
the properties of CSs.

In Fig.~\ref{fig:NM_charact} we display the posterior distributions 
for nine nuclear characteristics, see Eq.~\eqref{eq:Taylor_expansion}, 
at saturation density under scenarios B (left panels) and F (right panels). 
Note that the higher-order parameters $Z_{\rm sat}$ and $K_{\rm sym}$ are 
predictions of those lower-order parameters that uniquely determine the CDF. 
We conclude that the likelihood functions used for low-density matter 
constraints mainly affect the predictions for the distributions of 
nuclear characteristics.

In Fig.~\ref{fig:Heatmap}, we show the posterior correlation matrix for 
variations in nuclear characteristic parameters at saturation density, 
along with selected gross properties of CSs under the astrophysical 
scenario F, using two different likelihoods for low-density matter 
constraints. Scenario B yields similar findings. This figure also 
includes the dimensionless trace anomaly $\Delta$ and the squared 
sound speed $c_s^2$. The inferred nuclear characteristics are compiled 
in the Appendix, where we provide the parameters for each isospin sector 
up to fourth order in the expansion with respect to the density. 
In this work we use Pearson's correlation coefficient defined as
%--------------------------------
\begin{align}
r_{XY} = 
\frac{\text{cov}\,(X, Y)}{\sigma_X\,\sigma_Y}, \nonumber
\end{align}
%--------------------------------
where $\sigma_{X}$ is the standard deviation of $X$, 
$\text{cov}\,(X, Y)$ is the covariance between variables $X$ and $Y$, 
%--------------------------------
\begin{align}
\text{cov}\,(X, Y) = \frac{1}{n}\sum^n_i (X_i - \bar{X})\,(Y_i - \bar{Y}), \nonumber
\end{align}
%--------------------------------
with $n$ the number of samples and $\bar{X}$ denotes the mean of variable $X$.

From Figs.~\ref{fig:NM_charact} and~\ref{fig:Heatmap}, the following 
systematics are observed: 

(1). The posterior distributions for the three fundamental parameters, 
$M^\ast_{\rm D}$, $\rho_{\rm sat}$, and $E_{\rm sat}$, are quite similar 
to their prior distributions. This is especially evident for $\rho_{\rm sat}$ 
and $E_{\rm sat}$, suggesting that they are not important for the high-density 
properties of EOS. This conclusion is further supported by the correlation 
matrix shown in Fig.~\ref{fig:Heatmap}, where the absolute values of the 
Pearson correlation coefficients ($r$) between $E_{\rm sat}$ 
($\rho_{\rm sat}$) and the radii of selected stars, as well as the 
maximum mass $M_{\rm max}$, are relatively low, with $r$ less than 0.3.

(2). Another important observation is that in the UG2 approach, the Dirac 
effective mass $M^\ast_{\rm D}$ tends to be higher in both scenarios B 
and F, with a more significant increase for the PDFs in scenario B. This 
can likely be explained by the Hugenholtz-van-Hove theorem, applied at 
saturation density. It states that at zero temperature the Fermi energy 
equals the energy per particle at saturation~\citep{Boguta:1981,Boguta:1983}. 
Specifically, larger $M^\ast_{\rm D}$ which corresponds to small 
$\sigma$-meson field, requires a weaker $\omega$-meson field.  
Indeed, the Fermi energy is expressed as (disregard the rearrangement 
in the interaction)~\citep{Lijj:2018a,Sedrakian:2023}
%--------------------------------
\begin{align}
E_F = g_{\omega} \bar{\omega} + 
\sqrt{k^2_F + M^{\ast 2}_{\rm D}}. \nonumber
\end{align}
%--------------------------------
where $\bar{\omega}$ is the mean-field expectation value, and $k_F$ is the baryon Fermi momentum. At the saturation density, 
$E_F = E_{\rm sat}$, where $E_{\rm sat}$ is known to within approximately $5\%$. An increase in the Dirac mass $M^\ast_{\rm D}$ (corresponding to a reduction in the scalar potential) requires a smaller vector potential given by $g_{\omega} \bar{\omega}$. Consequently, this leads to a softer EOS at high densities, where the contribution from the 
$\omega$-meson dominates the behavior of the EOS. This results 
in a softer EOS as required by observational constraints in scenario B. Additionally, it is important to note a moderate 
anti-correlation ($r \approx -0.5$), shown in Fig.~\ref{fig:Heatmap}, between parameters $M^\ast_{\rm D}$ and $Q_{\rm{sat}}$, where the values of $Q_{\rm{sat}}$ derived from the UG2 approach are somewhat lower than those obtained using the Gaussian approach.

(3). The isoscalar incompressibility $K_{\rm{sat}}$ is highly sensitive to the likelihood functions and the observational constraints. In scenario B, the posterior distribution for $K_{\rm{sat}}$ is almost uniform across most of the distribution 
range when using the UG2 approach. In contrast, scenario F shows 
an exponential increase in the posterior, peaking near 
the upper limit of the uniform prior at $K_{\rm{sat}} = 310$~MeV. 
In both scenarios B and F, the UG2 approach produces wider 68.3\% and 95.4\% CI posterior ranges compared to the Gaussian approach (see the specific values in the Appendix).
Note that very large values of $K_{\rm{sat}} \sim 300$~MeV, are ruled out by the recent studies of the isoscalar GMR in nuclei~\citep{Litvinova:2022,Colo:2023}.

(4). In both scenarios B and F, the isocalar skewness values, $Q_{\rm{sat}}$, computed using the UG2 approach are somewhat lower compared to those obtained from the Gaussian approach. This adjustment is made to balance the effect of the higher $K_{\rm{sat}}$ values that the UG2 approach tends to favor. For example, in scenario B, the median value of $K_{\rm{sat}}$ increases by 12~MeV when transitioning from the Gaussian to the UG2 approaches, while the $Q_{\rm{sat}}$ values decrease by approximately 45~MeV. Note also that the coefficients in front of $K_{\rm{sat}}$ and $Q_{\rm{sat}}$ 
in Eq.~\eqref{eq:Taylor_expansion} differ by a factor of 1/3. 
We emphasize once more that $Q_{\rm{sat}}$ is the primary parameter dominating the maximum mass of CSs within the current CDF 
framework.

(5). The posteriors for the isovector parameters $J_{\rm sym}$ and $L_{\rm sym}$ still display Gaussian-like behaviors, even though the majority of the prior intervals are assumed to have uniform distributions (UG2). This behavior is due to the strong correlation between the two parameters when the $\chi$EFT constraint is applied. This is established in the correlation matrix shown in Fig.~\ref{fig:Heatmap}, where the Pearson correlation 
coefficient for the pair $(J_{\rm sym}, L_{\rm sym})$ is 
$r \approx 0.7$ in both approaches for the likelihoods.

(6). Similar to the counterbalancing behavior between the isoscalar parameters $K_{\rm{sat}}$ and $Q_{\rm{sat}}$, the slightly higher values of $L_{\rm sym}$ favored by the UG2 approach lead to an increase in the next-order isovector parameter, $K_{\rm{sym}}$, to keep the isovector sector of the EOS less sensitive to the choice of likelihood. 
Interestingly, we observe that $K_{\rm sym}$ also correlate strongly with $M_{\rm max}$.
However, it is important to note that within the current CDF framework, variations in $K_{\rm{sym}}$ primarily arise from adjustments to the lower-order isovector parameters $J_{\rm sym}$ and $L_{\rm sym}$.
Assessing the impacts of higher-order terms such as $K_{\text {sym }}$ would require extending the CDF framework to introduce additional degrees of freedom (DOFs). For example, one could generalize the density dependence of the $\rho$-meson coupling beyond the exponential form adopted in this work. Alternatively, incorporating an additional scalar-isovector $\delta$-meson could provide the necessary flexibility. 

Furthermore, we observe that the 95.4\% CI for $K_{\rm{sat}}$, obtained using the UG2 approach under scenario F, is shifted upward by approximately 20~MeV compared to the interval derived with a uniform prior. Although both intervals have comparable widths, this systematic shift toward higher values underscores important considerations for parameter estimation. In particular, the observed discrepancy highlights the need to carefully account for marginalization effects when employing uniform likelihood functions in Bayesian analyses. These results suggest that, unless the prior range is sufficiently broad, inadequate treatment of marginalization can introduce unintended biases in the posterior inference.
   
Finally, we return to the analysis of Fig.~\ref{fig:Heatmap}. The overall correlation structure among nuclear matter characteristics shows remarkable consistency between the two approaches. However, distinct differences emerge in specific parameter correlations: the absolute correlations between nuclear incompressibility $K_{\rm{sat}}$ and CS properties are noticeably enhanced when using the UG2 approach, compared to those derived from the Gaussian approach. In contrast, 
the correlation strengths involving the skewness parameter $Q_{\rm{sat}}$ exhibit significant attenuation under the UG2 approach. This behavior suggests a compensatory mechanism in the EOS parametrization, where these two parameters tend to offset each other's influence. In conclusion, our findings illustrate that the integrated nature of CS bulk properties inherently masks detailed information about individual nuclear parameters. Therefore, dedicated nuclear experiments specifically sensitive to $K_{\rm{sat}}$ and $L_{\rm sym}$ are essential for constraining their values more precisely.

The current CDF framework allows for the independent variation of saturation parameters up to 
$Q_{\rm sat}$ in the isoscalar sector and up to $L_{\rm sym}$ in the isovector sector~\citep{Lijj:2019a,Lijj:2023c}. Other types of CDFs with the same number of degrees of freedom (DOFs) as our model, for example, the FSU model~\citep{Horowitz:2001,Chen:2014}, are expected to yield qualitatively similar results. In contrast, models with fewer DOFs, such as those employed in Refs.~\citep {Malik:2022a,Zhu:2023,Beznogov:2023,Malik:2023}, often impose a strong correlation between 
$K_{\rm sat}$ and $Q_{\rm sat}$. As a result, the distribution of $K_{\rm sat}$ in those models is largely governed by that of $Q_{\rm sat}$, which itself is tightly constrained by the two-solar-mass limit on CS masses.
Conversely, models with more DOFs may exhibit compensating effects among parameters of different orders and across isospin multiplet sectors. This can lead to variations in the distribution of individual parameters. However, such differences do not alter the key conclusion that astrophysical data do not tightly constrain the lower-order saturation parameters.

\begingroup
\squeezetable
\begin{table*}[tb]
\caption{ 
Characteristic parameters of symmetric nuclear matter at 
saturation density from the posterior distributions for 
astrophysical scenarios Baseline, B and F with different 
likelihoods for low-density matter constraints. The 
superscripts and subscripts indicate the 68.3\% and 95.4\% 
(in parentheses) CI ranges.}
\setlength{\tabcolsep}{6.4pt}
\label{tab:NM_Posterior_BF}
\centering
\begin{tabular}{ccccccccc}
\hline\hline 
 \multirow{2}*{Par.} &\multirow{2}*{Unit}& 
 \multicolumn{3}{c}{Baseline}   & 
 \multicolumn{2}{c}{Scenario B} & 
 \multicolumn{2}{c}{Scenario F} \\
 \cline{3-5} \cline{6-7} \cline{8-9} 
 &  & Gaus.& UG1& UG2& Gaus.& UG2& Gaus.& UG2 \\   
\hline 
 $M_{\rm D}^\ast$            & $m_{\rm N}$     & 
 ${0.609}_{-0.043\,(0.091)}^{+0.041\,(0.081)}$ &
 ${0.610}_{-0.049\,(0.094)}^{+0.039\,(0.079)}$ &
 ${0.618}_{-0.071\,(0.113)}^{+0.056\,(0.080)}$ &
 ${0.614}_{-0.047\,(0.098)}^{+0.039\,(0.072)}$ & 
 ${0.630}_{-0.076\,(0.125)}^{+0.047\,(0.068)}$ &
 ${0.605}_{-0.043\,(0.091)}^{+0.035\,(0.070)}$ & 
 ${0.614}_{-0.067\,(0.110)}^{+0.053\,(0.083)}$ \\
 $\rho_{\rm sat}$            & fm$^{-3}$       &
 ${0.154}_{-0.005\,(0.009)}^{+0.005\,(0.010)}$ &
 ${0.154}_{-0.005\,(0.009)}^{+0.005\,(0.009)}$ &
 ${0.154}_{-0.007\,(0.011)}^{+0.006\,(0.009)}$ &
 ${0.154}_{-0.005\,(0.009)}^{+0.005\,(0.009)}$ & 
 ${0.155}_{-0.008\,(0.011)}^{+0.006\,(0.009)}$ &
 ${0.154}_{-0.005\,(0.009)}^{+0.005\,(0.009)}$ & 
 ${0.152}_{-0.007\,(0.009)}^{+0.007\,(0.011)}$ \\
\hline 
 $E_{\rm sat}$               & MeV             &
 ${-16.10}_{-0.20\,(0.40)}^{+0.20\,(0.40)}$    &
 ${-16.11}_{-0.20\,(0.40)}^{+0.20\,(0.40)}$    &
 ${-16.09}_{-0.30\,(0.41)}^{+0.28\,(0.39)}$    &
 ${-16.10}_{-0.20\,(0.40)}^{+0.20\,(0.40)}$    & 
 ${-16.08}_{-0.29\,(0.40)}^{+0.28\,(0.40)}$    &
 ${-16.10}_{-0.20\,(0.40)}^{+0.20\,(0.40)}$    & 
 ${-16.08}_{-0.29\,(0.42)}^{+0.27\,(0.38)}$    \\
 $K_{\rm sat}$               & MeV             &
 ${231.5}_{-35.5\,( 70.4)}^{+38.1\,(74.7)}$    &
 ${233.9}_{-39.1\,( 73.4)}^{+37.2\,(72.2)}$    &
 ${244.7}_{-60.2\,( 91.5)}^{+49.3\,(67.8)}$    &  
 ${231.4}_{-33.3\,( 64.0)}^{+34.4\,(68.7)}$    & 
 ${243.0}_{-54.2\,( 88.1)}^{+46.9\,(66.0)}$    &
 ${244.8}_{-34.7\,( 67.8)}^{+36.4\,(71.2)}$    & 
 ${270.7}_{-56.5\,(105.6)}^{+33.6\,(61.2)}$    \\ 
 $Q_{\rm sat}$               & MeV              &
 ${-361.5}_{-229.7\,(377.0)}^{+563.1\,(1245.5)}$&
 ${-354.7}_{-228.6\,(378.0)}^{+569.4\,(1322.8)}$&
 ${-393.7}_{-222.4\,(413.9)}^{+548.4\,(1418.2)}$& 
 ${-391.0}_{-170.2\,(323.3)}^{+333.6\,( 872.8)}$& 
 ${-436.0}_{-183.7\,(434.1)}^{+304.2\,( 909.1)}$&
 ${-164.7}_{-280.1\,(455.4)}^{+548.9\,(1161.9)}$& 
 ${-298.5}_{-209.1\,(483.5)}^{+515.8\,(1266.8)}$\\
 $Z_{\rm sat}$               & MeV             &
 ${4866}_{-2377\,(4978)}^{+1704\,(4920)}$      &
 ${4829}_{-2482\,(4970)}^{+1818\,(5241)}$      &
 ${4294}_{-3351\,(5020)}^{+2801\,(7626)}$      & 
 ${4931}_{-2330\,(4721)}^{+1595\,(4862)}$      & 
 ${4386}_{-3303\,(4920)}^{+2693\,(8369)}$      &
 ${4641}_{-2508\,(5071)}^{+1663\,(5497)}$      & 
 ${3367}_{-3145\,(4818)}^{+2962\,(8739)}$      \\ 
\hline         
 $J_{\rm sym}$               & MeV             &
 ${31.20}_{-0.87\,(1.69)}^{+0.88\,(1.80)}$     &
 ${31.18}_{-0.84\,(1.79)}^{+0.97\,(1.95)}$     &
 ${31.06}_{-1.38\,(2.32)}^{+1.68\,(3.21)}$     & 
 ${31.24}_{-0.86\,(1.68)}^{+0.88\,(1.78)}$     & 
 ${31.27}_{-1.45\,(2.44)}^{+1.66\,(3.14)}$     &
 ${31.20}_{-0.88\,(1.72)}^{+0.89\,(1.82)}$     & 
 ${31.31}_{-1.46\,(2.47)}^{+1.65\,(3.12)}$     \\ 
 $L_{\rm sym}$               & MeV             &
 ${46.28}_{-5.91\,(11.44)}^{+6.33\,(13.13)}$   &
 ${46.89}_{-6.75\,(12.95)}^{+6.18\,(12.33)}$   &
 ${47.82}_{-9.58\,(17.84)}^{+10.05\,(18.74)}$  &
 ${46.73}_{-5.67\,(11.35)}^{+5.94\,(12.10)}$   & 
 ${48.63}_{-8.84\,(16.79)}^{+9.46\,(17.39)}$   &
 ${47.86}_{-6.27\,(12.15)}^{+6.46\,(13.08)}$   & 
 ${51.73}_{-9.04\,(18.35)}^{+9.06\,(16.72)}$   \\
 $K_{\rm sym}$               & MeV             &
 ${-103.1}_{-18.5\,(31.5)}^{+27.8\,(63.3)}$    &
 ${-103.1}_{-19.0\,(32.2)}^{+29.6\,(70.0)}$    &
 ${-104.0}_{-20.8\,(34.8)}^{+34.6\,(91.9)}$    & 
 ${-106.0}_{-14.7\,(25.8)}^{+21.0\,(50.5)}$    & 
 ${-110.0}_{-15.8\,(27.7)}^{+27.9\,(73.1)}$    &
 ${- 94.2}_{-18.2\,(31.1)}^{+26.6\,(61.8)}$    & 
 ${-101.8}_{-17.7\,(30.9)}^{+30.2\,(83.4)}$    \\
 $Q_{\rm sym}$               & MeV             &
 ${792.8}_{-173.0\,(351.8)}^{+155.9\,(289.6)}$ &
 ${776.2}_{-166.5\,(336.2)}^{+172.6\,(313.7)}$ &
 ${715.4}_{-215.5\,(406.7)}^{+233.6\,(423.6)}$ & 
 ${776.1}_{-159.1\,(323.0)}^{+155.5\,(295.8)}$ & 
 ${702.6}_{-194.7\,(366.3)}^{+212.5\,(409.9)}$ &
 ${765.3}_{-177.0\,(358.4)}^{+166.5\,(299.3)}$ & 
 ${624.8}_{-180.4\,(361.6)}^{+225.7\,(440.4)}$ \\
 $Z_{\rm sym}$               & MeV             &
 ${-5369}_{-2398\,(5401)}^{+1660\,(2840)}$     &
 ${-5302}_{-2572\,(5952)}^{+1669\,(2763)}$     &
 ${-4887}_{-3149\,(7570)}^{+2004\,(3066)}$     & 
 ${-5079}_{-1988\,(4606)}^{+1435\,(2523)}$     & 
 ${-4402}_{-2589\,(6298)}^{+1692\,(2587)}$     &
 ${-5686}_{-2596\,(5565)}^{+1820\,(3095)}$     & 
 ${-4401}_{-3007\,(7611)}^{+1705\,(2630)}$     \\
\hline\hline 
\end{tabular}
\end{table*}
\endgroup

%----------------------------------------------------------
\section{Conclusions}
\label{sec:Conclusions}
%----------------------------------------------------------
In this work, we performed a systematic comparison of Bayesian 
inference outcomes for CDF-type EOS of dense matter, focusing on 
the impacts of likelihood functions between Gaussian and uniform 
distributions when incorporating low-density matter constraints. 
While both likelihood formulations are routinely utilized in 
Bayesian analyses, we introduce critical modifications to ensure 
methodological equivalence: the uniform likelihood is augmented 
with a normalization factor and a Gaussian-inspired tail.
In both approaches, the EOS models have been built with identical
priors for the CDF parameters, and employ the same methods to 
construct the likelihoods for observational constraints.

In our analysis, we have assessed how the two inferences
differ by testing the gross properties of CSs and dense matter, 
and the correlations among them, under three astrophysical 
scenarios providing looser and tighter constraints, respectively.   
We have found that the two approaches lead to very
similar posterior results for the gross properties of CSs 
(e.g., maximum mass, radii, and tidal deformabilities)
and dense matter (e.g., EOS, sound speed, and trace
anomaly) with the 95.4\% CI essentially overlapping.
The difference in particle composition is statistically 
negligible and can be safely neglected without noticeable 
effect.

Notable discrepancies emerge in the nuclear matter characteristics 
at saturation density. Parameters such as the saturation density 
$\rho_{\rm sat}$ and energy per particle $E_{\rm sat}$, which are 
largely independent of observational constraints, retain prior-dominated 
distributions. In contrast, the incompressibility modulus $K_{\rm{sat}}$ 
shows high sensitivity to both its likelihood function and the 
observational data, while the isoscalar skewness $Q_{\rm{sat}}$ 
adjusts accordingly to preserve the consistency of the predictions 
of the EOS. In the isovector sector, the parameters $J_{\rm sym}$ 
and $L_{\rm sym}$ maintain a strong mutual correlation, even though 
their individual distributions vary depending on the choice of 
likelihood functions. These findings highlight the fact that the 
integrated nature of CS gross properties tends to obscure direct 
information about individual nuclear parameters close to saturation. 

\section*{Acknowledgments}
J.-J.L. acknowledges the support of the National Natural 
Science Foundation of China under Grants No. 12105232 and 
No. 12475150. 
A.S. acknowledges funding by the Deutsche Forschungsgemeinschaft 
Grant No. SE 1836/6-1 and the Polish NCN Grant No. 
2023/51/B/ST9/02798.

%----------------------------------------------------------
\appendix*
\section{Key quantities of nuclear matter and compact stars}
\label{appendix}
%----------------------------------------------------------
In this appendix, we present the characteristic parameters of 
nuclear matter at saturation density (in Table~\ref{tab:NM_Posterior_BF}) 
and key gross quantities of CSs (in Table~\ref{tab:NS_Posterior_BF}) 
predicted by CDFs for the three astrophysical scenarios, 
assuming for the low-density matter constraints a likelihood 
of Gaussian distribution (Gaus.) or a combination of uniform 
and Gaussian distributions (UG1 or UG2).

\begingroup
\squeezetable
\begin{table*}[tb]
\caption{
Key quantities of compact stars from the posterior distributions 
for astrophysical scenarios Baseline, B and F with different 
likelihoods for low-density matter constraints: radii, tidal 
deformabities, central baryonic densities, energy densities, 
pressures, sound speeds, and trace anomalies for 
$1.0,\,1.4,\,2.0\,M_{\odot}$ and the maximum-mass stars.
The superscripts and subscripts indicate the 95.4\% CI ranges.}
\setlength{\tabcolsep}{10.0pt}
\label{tab:NS_Posterior_BF}
\centering
\begin{tabular}{ccccccccc} 
\hline\hline 
 \multirow{2}*{Par.} &\multirow{2}*{Unit}& 
 \multicolumn{3}{c}{Baseline}   & 
 \multicolumn{2}{c}{Scenario B} & 
 \multicolumn{2}{c}{Scenario F} \\
 \cline{3-5} \cline{6-7} \cline{8-9} 
 &  & Gaus.& UG1& UG2& Gaus.& UG2& Gaus.& UG2 \\   
\hline 
     $R_{1.0}$                  & km                           &
 ${12.47}_{-0.71}^{+0.58}$      & ${12.51}_{-0.76}^{+0.58}$    &
 ${12.56}_{-0.89}^{+0.63}$      & ${12.44}_{-0.43}^{+0.38}$    & 
 ${12.50}_{-0.49}^{+0.42}$      & ${12.69}_{-0.47}^{+0.43}$    & 
 ${12.81}_{-0.56}^{+0.43}$      \\ 
     $\Lambda_{1.0}$            &                              &
 ${3162}_{-995}^{+1126}$        & ${3222}_{-1059}^{+1151}$     &
 ${3307}_{-1189}^{+1213}$       & ${3114}_{-619}^{+667}$       &
 ${3184}_{-700}^{+693}$         & ${3554}_{-764}^{+858}$       &
 ${3706}_{-840}^{+883}$        \\ 
     $\rho_{1.0}$               & fm$^{-3}$                    &
 ${0.339}_{-0.059}^{+0.076}$    & ${0.336}_{-0.060}^{+0.078}$  &
 ${0.334}_{-0.061}^{+0.082}$    & ${0.342}_{-0.041}^{+0.044}$  &    
 ${0.341}_{-0.038}^{+0.046}$    & ${0.316}_{-0.040}^{+0.047}$  &
 ${0.315}_{-0.041}^{+0.045}$   \\
     $P_{1.0}$                  & MeV/fm$^{3}$                 &
 ${29.89}_{-6.66}^{+10.01}$     & ${29.50}_{-6.65}^{+10.39}$   &
 ${29.07}_{-6.61}^{+11.42}$     & ${30.27}_{-4.52}^{+5.69}$    &   
 ${29.97}_{-4.32}^{+6.13}$      & ${27.19}_{-4.43}^{+5.69}$    &
 ${26.72}_{-4.32}^{+5.66}$     \\ 
     $\varepsilon_{1.0}$        & MeV/fm$^{3}$                 &
 ${337.30}_{-61.49}^{+78.67}$   & ${334.43}_{-61.96}^{+80.99}$ &
 ${333.22}_{-62.90}^{+84.76}$   & ${340.87}_{-42.40}^{+45.95}$ &
 ${340.34}_{-40.02}^{+48.04}$   & ${313.91}_{-41.88}^{+48.94}$ &
 ${313.33}_{-42.73}^{+46.76}$  \\ 
     $c_{s, 1.0}^2$             &                              &
 ${0.267}_{-0.043}^{+0.032}$    & ${0.266}_{-0.043}^{+0.032}$  &
 ${0.260}_{-0.045}^{+0.041}$    & ${0.263}_{-0.037}^{+0.033}$  &    
 ${0.255}_{-0.036}^{+0.042}$    & ${0.269}_{-0.044}^{+0.030}$  &
 ${0.251}_{-0.039}^{+0.047}$   \\ 
     $\Delta_{1.0}$             &                              &
 ${0.245}_{-0.008}^{+0.005}$    & ${0.245}_{-0.008}^{+0.005}$  &
 ${0.246}_{-0.010}^{+0.006}$    & ${0.244}_{-0.005}^{+0.004}$  &       
 ${0.245}_{-0.006}^{+0.004}$    & ${0.247}_{-0.005}^{+0.004}$  &
 ${0.248}_{-0.006}^{+0.004}$   \\ 
\hline
     $R_{1.4}$                  & km                           &
 ${12.51}_{-0.84}^{+0.77}$      & ${12.55}_{-0.88}^{+0.78}$    &
 ${12.58}_{-0.95}^{+0.80}$      & ${12.47}_{-0.50}^{+0.48}$    &    
 ${12.49}_{-0.52}^{+0.48}$      & ${12.79}_{-0.56}^{+0.55}$    &
 ${12.84}_{-0.57}^{+0.56}$     \\   
     $\Lambda_{1.4}$            &                              &
 ${483}_{-189}^{+271}$          & ${493}_{-198}^{+283}$        &
 ${497}_{-206}^{+295}$          & ${472}_{-121}^{+163}$        &      
 ${474}_{-124}^{+153}$          & ${571}_{-162}^{+207}$        &
 ${574}_{-156}^{+218}$         \\   
     $\rho_{1.4}$               & fm$^{-3}$                    &
 ${0.420}_{-0.092}^{+0.113}$    & ${0.416}_{-0.093}^{+0.116}$  &
 ${0.419}_{-0.098}^{+0.114}$    & ${0.425}_{-0.067}^{+0.070}$  &    
 ${0.429}_{-0.066}^{+0.067}$    & ${0.384}_{-0.062}^{+0.078}$  & 
 ${0.391}_{-0.069}^{+0.069}$   \\  
     $P_{1.4}$                  & MeV/fm$^{3}$                   &
 ${58.979}_{-17.626}^{+26.249}$ & ${58.189}_{-17.601}^{+26.827}$ &
 ${58.260}_{-18.156}^{+27.410}$ & ${60.042}_{-12.859}^{+15.313}$ &     
 ${60.402}_{-12.345}^{+15.269}$ & ${51.927}_{-11.458}^{+15.549}$ &
 ${52.511}_{-12.267}^{+14.136}$ \\  
     $\varepsilon_{1.4}$        & MeV/fm$^{3}$                   &
 ${428.05}_{-99.64}^{+124.05}$  & ${424.25}_{-100.70}^{+126.67}$ &
 ${427.55}_{-105.79}^{+124.68}$ & ${433.76}_{-73.38}^{+76.56}$   &     
 ${438.23}_{-71.62}^{+74.32}$   & ${389.96}_{-66.82}^{+85.26}$   &
 ${397.99}_{-75.01}^{+75.24}$   \\  
     $c_{s, 1.4}^2$             &                                &
 ${0.378}_{-0.052}^{+0.029}$    & ${0.377}_{-0.052}^{+0.029}$    &
 ${0.371}_{-0.058}^{+0.040}$    & ${0.374}_{-0.046}^{+0.030}$    &     
 ${0.368}_{-0.050}^{+0.039}$    & ${0.376}_{-0.054}^{+0.026}$    &
 ${0.359}_{-0.057}^{+0.042}$    \\
     $\Delta_{1.4}$             &                                &
 ${0.196}_{-0.017}^{+0.012}$    & ${0.196}_{-0.018}^{+0.012}$    &
 ${0.197}_{-0.020}^{+0.012}$    & ${0.195}_{-0.010}^{+0.008}$    &    
 ${0.195}_{-0.011}^{+0.008}$    & ${0.200}_{-0.010}^{+0.008}$    &
 ${0.202}_{-0.011}^{+0.008}$    \\ 
\hline    
     $R_{2.0}$                  & km                             &
 ${12.15}_{-1.47}^{+1.26}$      & ${12.19}_{-1.50}^{+1.28}$      &
 ${12.12}_{-1.42}^{+1.38}$      & ${12.04}_{-1.03}^{+0.91}$      &     
 ${11.98}_{-1.00}^{+0.90}$      & ${12.60}_{-1.14}^{+0.88}$      &
 ${12.49}_{-1.08}^{+1.00}$      \\ 
     $\Lambda_{2.0}$            &                                &
 ${36.0}_{-25.1}^{+47.9}$       & ${37.0}_{-26.1}^{+50.6}$       &
 ${34.7}_{-24.0}^{+54.0}$       & ${33.2}_{-19.1}^{+30.6}$       &    
 ${31.0}_{-17.5}^{+29.3}$       & ${48.9}_{-28.6}^{+38.6}$       &
 ${44.2}_{-25.4}^{+43.4}$       \\
     $\rho_{2.0}$               & fm$^{-3}$                      &
 ${0.615}_{-0.194}^{+0.381}$    & ${0.608}_{-0.194}^{+0.384}$    &
 ${0.629}_{-0.215}^{+0.357}$    & ${0.636}_{-0.162}^{+0.273}$    &     
 ${0.657}_{-0.168}^{+0.271}$    & ${0.539}_{-0.126}^{+0.254}$    &  
 ${0.572}_{-0.156}^{+0.263}$    \\
     $P_{2.0}$                  & MeV/fm$^{3}$                   &
 ${182.14}_{-86.35}^{+279.85}$  & ${178.31}_{-85.35}^{+280.73}$  &
 ${188.22}_{-95.63}^{+265.61}$  & ${193.66}_{-76.50}^{+185.95}$  &     
 ${204.72}_{-81.94}^{+190.34}$  & ${144.69}_{-51.71}^{+143.77}$  &  
 ${158.67}_{-65.25}^{+154.34}$  \\
     $\varepsilon_{2.0}$        & MeV/fm$^{3}$                   &
 ${677.39}_{-236.40}^{+544.04}$ & ${667.93}_{-235.34}^{+547.58}$ &
 ${696.14}_{-263.46}^{+513.13}$ & ${704.78}_{-201.48}^{+385.85}$ &     
 ${732.56}_{-212.25}^{+388.25}$ & ${582.67}_{-150.38}^{+339.60}$ &
 ${624.62}_{-189.97}^{+359.34}$ \\
     $c_{s, 2.0}^2$             &                                &
 ${0.573}_{-0.044}^{+0.087}$    & ${0.571}_{-0.044}^{+0.089}$    &
 ${0.570}_{-0.054}^{+0.092}$    & ${0.577}_{-0.036}^{+0.057}$    &     
 ${0.578}_{-0.046}^{+0.063}$    & ${0.557}_{-0.036}^{+0.045}$    &
 ${0.552}_{-0.053}^{+0.054}$    \\ 
     $\Delta_{2.0}$             &                                & 
 ${0.065}_{-0.109}^{+0.052}$    & ${0.066}_{-0.111}^{+0.052}$    &
 ${0.063}_{-0.106}^{+0.057}$    & ${0.059}_{-0.074}^{+0.042}$    &     
 ${0.054}_{-0.075}^{+0.044}$    & ${0.085}_{-0.065}^{+0.033}$    & 
 ${0.080}_{-0.066}^{+0.039}$    \\ 
\hline    
     $M_{\rm max}$              & $M_{\odot}$                    &
 ${2.21}_{-0.23}^{+0.32}$       & ${2.22}_{-0.23}^{+0.33}$       &
 ${2.20}_{-0.21}^{+0.35}$       & ${2.20}_{-0.17}^{+0.23}$       &      
 ${2.18}_{-0.16}^{+0.22}$       & ${2.32}_{-0.24}^{+0.24}$       &
 ${2.26}_{-0.22}^{+0.29}$       \\ 
     $R_{M_{\rm max}}$          & km                             &
 ${11.05}_{-0.88}^{+1.18}$      & ${11.09}_{-0.91}^{+1.22}$      &
 ${11.04}_{-0.86}^{+1.27}$      & ${11.00}_{-0.58}^{+0.80}$      &     
 ${10.95}_{-0.53}^{+0.76}$      & ${11.44}_{-0.78}^{+0.88}$      &
 ${11.32}_{-0.65}^{+0.97}$      \\ 
     $\Lambda_{M_{\rm max}}$    &                                &
 ${5.97}_{-1.04}^{+2.02}$       & ${5.95}_{-1.03}^{+2.02}$       &
 ${6.08}_{-1.10}^{+2.52}$       & ${6.04}_{-0.93}^{+1.78}$       &      
 ${6.19}_{-1.01}^{+2.19}$       & ${5.64}_{-0.73}^{+2.18}$       & 
 ${6.02}_{-1.08}^{+3.36}$       \\ 
     $\rho_{M_{\rm max}}$       & fm$^{-3}$                      &
 ${0.985}_{-0.205}^{+0.201}$    & ${0.979}_{-0.208}^{+0.206}$    &
 ${0.994}_{-0.222}^{+0.189}$    & ${0.996}_{-0.152}^{+0.136}$    &       
 ${1.010}_{-0.150}^{+0.125}$    & ${0.912}_{-0.142}^{+0.169}$    &  
 ${0.942}_{-0.169}^{+0.146}$    \\ 
     $P_{M_{\rm max}}$          & MeV/fm$^{3}$                   &
 ${583.58}_{-89.44}^{+100.49}$  & ${580.05}_{-90.82}^{+103.00}$  &
 ${578.91}_{-93.59}^{+110.00}$  & ${588.95}_{-62.70}^{+62.09}$   &     
 ${590.41}_{-61.61}^{+68.87}$   & ${551.63}_{-63.29}^{+69.28}$   & 
 ${551.81}_{-69.80}^{+71.91}$   \\  
 $\varepsilon_{M_{\rm max}}$     &  MeV/fm$^{3}$                   & 
 ${1284.91}_{-269.12}^{+263.71}$ & ${1276.17}_{-273.02}^{+270.00}$ &
 ${1295.57}_{-291.27}^{+250.76}$ & ${1298.90}_{-199.98}^{+177.49}$ &  
 ${1317.69}_{-197.72}^{+165.20}$ & ${1189.94}_{-188.08}^{+219.07}$ &  
 ${1227.19}_{-221.17}^{+188.74}$\\  
     $c_{s, M_{\rm max}}^2$     &                                &
 ${0.728}_{-0.055}^{+0.039}$    & ${0.729}_{-0.055}^{+0.038}$     &
 ${0.726}_{-0.067}^{+0.040}$    & ${0.726}_{-0.049}^{+0.032}$     &      
 ${0.723}_{-0.061}^{+0.033}$    & ${0.741}_{-0.063}^{+0.028}$     &  
 ${0.730}_{-0.096}^{+0.038}$    \\ 
     $\Delta_{M_{\rm max}}$     &                                &
 ${-0.125}_{-0.031}^{+0.043}$   & ${-0.126}_{-0.031}^{+0.043}$   &
 ${-0.123}_{-0.032}^{+0.051}$   & ${-0.123}_{-0.027}^{+0.038}$   &     
 ${-0.120}_{-0.028}^{+0.045}$   & ${-0.134}_{-0.023}^{+0.049}$   & 
 ${-0.125}_{-0.031}^{+0.067}$   \\ 
\hline\hline 
\end{tabular}
\end{table*}
\endgroup

%\newpage
%\bibliography{Bayesian_refs}

%

\end{document}